\documentclass[twocolumn,english,trans]{IEEEtran}
\usepackage{amsfonts}
\usepackage{float}
\usepackage{amssymb}
\usepackage[T1]{fontenc}
\usepackage{amsthm}
\usepackage{units}
\usepackage{changebar}
\usepackage{color}
\usepackage{subfigure}
\newtheorem{prop}{Proposition}

\usepackage{cite}

\ifCLASSINFOpdf
  \usepackage[pdftex]{graphicx}
   \graphicspath{{../pdf/}{../jpeg/}}
  \DeclareGraphicsExtensions{.pdf,.jpeg,.png}
\else
  \usepackage[dvips]{graphicx}
  \graphicspath{{../eps/}}
  \DeclareGraphicsExtensions{.eps}
\fi
\usepackage[cmex10]{amsmath}
\allowdisplaybreaks

\usepackage{subfigure}

\begin{document}
\title{Phase-Rotation-Aided Relay Selection in Two-Way Decode-and-Forward Relay Networks}
\author{Ruohan~Cao,~Hui~Gao,~\textit{Member,~IEEE,}~Tiejun~Lv,~\textit{Senior Member,~IEEE,}
~Shaoshi~Yang,~\textit{Member,~IEEE}~and~Shanguo Huang%
\thanks{Copyright (c) 2015 IEEE. Personal use of this material is permitted. However, permission to use this material for any other purposes must be obtained from the IEEE by sending a request to pubs-permissions@ieee.org.}
\thanks{Manuscript received September 20, 2014; revised March 20, 2015;
accepted May 30, 2015. The editor coordinating the reviewer of this paper and
approving it for publication was Walaa Hamouda. This work is supported by the National Natural Science Foundation of
China (NSFC) (Grant No. 61271188, 61401041, and 61331008), the National High Technology Research and Development Program of China (863 Program) (Grant No. 2015AA01A706), the Fundamental Research Funds for the Central Universities (Grant No. 2014RC0106), and Beijing Municipal Science and Technology Commission Research Fund Project (Grant No. D151100000115002).}
\thanks{R.~Cao and S.~Huang are with the Institute of Information Photonics and Optical Communications, Beijing University of Posts and Telecommunications (BUPT), Beijing 100876, China,
(e-mail: \{caoruohan, shghuang\}@bupt.edu.cn). R.~Cao is also with the School of Information and Communication Engineering,
BUPT. }
\thanks{H.~Gao and T.~Lv are with the School of Information and Communication Engineering,
BUPT, Beijing 100876, China,
(e-mail: \{huigao, lvtiejun\}@bupt.edu.cn).}
\thanks{S. Yang is with the School of Electronics and
Computer Science, University of Southampton, SO17 1BJ Southampton, U.K. (e-mail: sy7g09@ecs.soton.ac.uk).}


}

\maketitle
\begin{abstract}
This paper proposes a relay selection scheme that aims to improve the end-to-end symbol error rate (SER) performance of a two-way relay network (TWRN). The TWRN consists of two single-antenna sources and multiple relays employing decode-and-forward (DF) protocol. It is shown that the SER performance is determined by the minimum decision distance (DD) observed in the TWRN. However, the minimum DD is likely to be made arbitrarily small by channel fading. To tackle this problem, a phase rotation (PR) aided relay selection (RS) scheme is proposed to enlarge the minium DD, which in turn improves the SER performance. The proposed PR based scheme rotates the phases of the transmitted symbols of one source and of the selected relay according to the  channel state information, aiming for increasing all DDs to be above a desired bound. The lower bound is further optimized by using a MaxMin-RS criterion associated with the channel gains. It is demonstrated that the PR aided MaxMin-RS approach achieves full diversity gain and an improved array gain. Furthermore, compared with the existing DF based schemes, the proposed scheme allows more flexible relay antenna configurations.
\end{abstract}
\begin{IEEEkeywords}
Decode-and-forward, beamforming, relay selection, network coding, MIMO, full diversity.
\end{IEEEkeywords}

%
\IEEEpeerreviewmaketitle

%
%
%
%

\section{Introduction}

\IEEEPARstart{T}wo-way relaying (TWR) is a promising technique to
improve the coverage and connectivity of relay aided
networks \cite{larsson2006coded,kim2008achievable,kim2008performance}.
In a typical TWR channel (TWRC) \cite{BRankov}, \cite{EVan}, two source nodes exchange information simultaneously with the
aid of a relay node. Assuming the absence of a direct link between the two source
nodes, communication takes place in two stages: the multiple access (MA) stage and the broadcast (BC) stage. During the MA stage, both source nodes transmit their individual signals to
the relay node simultaneously. Then, the relay node broadcasts the processed signals to the source
nodes in the BC stage. It is also assumed that the TWR
schemes are assisted by the network coding in analog or digital domain
\cite{zhang2006hot}, \cite{AnalogISIST}. In fading wireless channels,
multiple relay antennas can bring diversity gain to the TWR system\cite{gunduz2008mimo}. This paper focuses on  achieving diversity in the network coding aided TWR systems composed of multiple relay antennas and two single-antenna sources. The performance metric considered is the symbol error rate (SER), based on which the achievable diversity gain is derived.

In terms of the symbol error rate (SER) performance, diversity techniques have been studied extensively for TWR systems. These existing works generally obtain full diversity gain by processing signal(s) from either all relay antennas or a single one selected relay antenna. To be more specific,  in \cite{eslamifarmax}, \cite{caospnc}, the received signals from all relay antennas are combined, which allows both diversity gain and array gain to be achieved. The signal combining operation employed in \cite{eslamifarmax}, \cite{caospnc} obstructs its application in the distributed scenario, where the multiple relay antennas are  distributed among several single-antenna relays. This limitation is removed by the distributed-antenna space-time coding (DSTC) based scheme \cite{cui2009distributed} and a variety of antenna selection (AS) based schemes \cite{jing2009relay, AF-SER, zhou2010decode}. In particular, the DSTC is implemented in \cite{cui2009distributed} with the cooperation among all antennas. The performance achieved by the scheme of \cite{cui2009distributed} is further enhanced by the AS based techniques \cite{jing2009relay, AF-SER, zhou2010decode}. All these AS based schemes \cite{jing2009relay, AF-SER, zhou2010decode} use the MaxMin criterion, which maximizes the quality indicator of the worst link originating from the selected relay to the two source nodes. The MaxMin criterion
is able to obtain full diversity gain by employing the amplify-and-forward (AF) relaying protocol to the selected antenna \cite{jing2009relay},\cite{AF-SER}. The decode-and-forward (DF) protocol is considered by a opportunistic two-way relaying (O-TR) scheme \cite{zhou2010decode}, which also invokes MaxMin criterion and the resultant SER performance exhibits full diversity order, although the scheme is aided by perfect error-correction-code (ECC). In other words, the DF protocol in the O-TR scheme is carried out on the codeword by codeword basis. However, it is observed that in the absence of ECC, the MaxMin criterion aided DF protocol no longer guarantees full diversity gain.

Against this background, in this paper we aim to tackle the problem of full-diversity guaranteed uncoded transmission, where the DF protocol is implemented on the symbol by symbol basis. We focus on the scenario with $K$ ($K\geq1$) relays each equipped with an arbitrary number of antennas. Hence, the existing scenarios considered in \cite{cui2009distributed, jing2009relay, AF-SER, zhou2010decode, krikidis2010relay, eslamifarmax,caospnc} are its special cases. Intuitively, a natural transmission approach is the straightforward extension of the previous AS based schemes, which always select a single antenna despite the distribution of the total relaying antennas mounted on $K$ relay(s). By comparison, instead of simply extending these works, in this paper we propose a relay selection (RS) based scheme, which not only gleans full diversity gain from relay selection, but also improves array array power gain by combining signals from multiple antennas of the selected relay. Therefore, the proposed RS scheme enjoys the advantage of array array power gain over the existing AS based schemes.

More specifically, the key enabler of the proposed RS scheme is a phase rotation (PR) strategy. In general, the PR strategy is executed symbol by symbol, which intends to remove the randomness of the phase metric of TWRC. This phase metric is termed effective angle (EA). In our previous work \cite{caospnc}, the randomness of EA is shown to degrade the achievable diversity gain in the MA stage. The proposed scheme in \cite{caospnc} is  applicable to either the single multi-antenna relay or the selected single-antenna relay, and it employs a sub-optimal detector specific for the MPSK modulation. By contrast, in this paper we consider a more flexible antenna distribution and tackle the EA randomness problem under arbitrary modulation. Throughout this paper, the selected relay employs the optimal maximum likelihood (ML) detector, which is in principle more fundamental than the sub-optimal detector proposed in \cite{caospnc}. Furthermore, we point out that in the more general scenario considered, the randomness of EA impose an impact on the achievable performance in both the MA and BC stages. Correspondingly, as the derivatives of the PR strategy, the PR preprocessing and PR beamforming operations are conceived for the MA and BC stages, respectively. By removing the randomness of EA in both stages, the PR operations shape the distribution of the decision-distances (DDs), i.e., the Euclidean distances between all desired received signal points, as observed by receivers. It is demonstrated that the stochastic property of reshaped DDs allows the MaxMin-RS criterion to attain full diversity when the DF protocol is executed symbol by symbol.

Notice that some existing approaches also process phase of symbols transmitted in one-way relay system which seems a little bit similar to the proposed PR strategy \cite{yang2010distributed,yang2011RF,yang2011allerton}. However, the full diversity gain achieved by these approaches is measured by outage capacity. This paper focuses on the influence of PR strategy on SER performance. All the diversity analysis provided by this paper are in the sense of SER performance rather than outage capacity. This is striking difference between our work and the other existing paper \cite{yang2010distributed,yang2011RF,yang2011allerton}. Generally, the diversity analysis method and its result derived from outage capacity are not equivalent to that from SER performance. For example, it is demonstrated that according to outage capacity, MaxMin-RS without any preprocessing is capable of obtaining full diversity gain \cite{krikidis2010relay}. However, our previous work has shown SER performance does not exhibit any diversity gain by applying MaxMin-RS straightforwardly \cite{caospnc}. This paper indeed provides insights into how PR strategy enhance the capability of MaxMin-RS on achieving diversity gains in SER performance.

In particular, we provide theoretical analysis to confirm the impact of pre-canceling the randomness of the EA on achievable SER performance. The full diversity order property and the achievable SER performance of the proposed scheme are explicitly analyzed. Our analysis indicates that by exploiting the information embedded in the TWRC's phase metric, the SER performance can be improved by using the MaxMin-RS approach in uncoded DF based relay networks.

The rest of this paper is organized as follows. In Section II, the system model is described. In Section III, we present the proposed PR-MaxMin-RS scheme. In Section IV, we provide more details of the PR based approach, and derive the SER expression under certain system configurations. In Section V, numerical results are offered to confirm the advantages of the proposed scheme and to validate the theoretical analysis of the diversity order. Finally, we conclude the paper in Section VI.

\emph{Notation}: For a matrix, ${\left(\cdot\right)^{H}}$,${\left(\cdot\right)^{T}}$,
and ${\left\Vert \cdot \right\Vert}$
represent the conjugate transpose, transpose,
and Frobenius norm of the matrix, respectively. For a complex-valued variable, $\Re\left(\cdot\right)$, $\Im\left(\cdot\right)$, $| \cdot |$, and ${\left(\cdot\right)^{*}}$
denote the real part, the imaginary part, the absolute value and the conjugate of the complex-valued variable, respectively. For a set, $| \cdot |$ represents the size of this set. $\mathcal{CN}\left(\mathbf{0},\mathbf{K}\right)$ denotes a circularly symmetric Gaussian random vector with mean $\mathbf{0}$ and covariance matrix $\mathbf{K}$. $\mathcal{N}\left(0,1\right)$ denotes a standard Gaussian distribution. $\angle\cdot$ stands for the angle of a complex-valued number. $\mathbf{0}_{M\times N}$ denotes the $M\times N$ matrix whose all entries are
zero, and $\mathbf{I}_{M\times N}$ denotes the $M\times N$ unit matrix. ${\mathbb{E}}\left\{ \cdot\right\}$ represents the expectation operation with respect to its argument. $Q\left(\cdot\right)$ is the tail of the probability density function (PDF) of a standard Gaussian random variable, i.e., $Q\left(x\right)=\frac{1}{\pi}\int_{0}^{\frac{\pi}{2}}\exp\left(-\frac{x^{2}}{2\sin^{2}\theta}\right)d\theta$.

\section{SYSTEM MODEL}
\subsection{Channel Model}
We consider the DF based TWRC shown in Fig.1, where two single-antenna source nodes ${\rm{S}}_{i}$ $\left(i=1,2\right)$
exchange messages with the aid of $K$ half-duplex relay nodes denoted as $\left\{ \textrm{R}_{k}\left|k=1,\ldots,K\right.\right\}$. The relay node ${{\rm{R}}_k}$ is equipped with
${L_k}$ antennas, where ${L_k} \ge 1$, for $k=1,\ldots,K$. Let $ L=\sum\limits_{k = 1}^K {{L_k}}$. We assume that there is no direct link connecting the source nodes, while the links between the two source nodes and relays suffer small-scale fading. Specifically, the links from ${\rm{S}}_{1}$ and ${\rm{S}}_{2}$ to ${{\rm{R}}_k}$ are characterized by the $L_k\times1$
channel vectors $\mathbf{h}_{k}$ and $\mathbf{g}_{k}$, respectively. The elements $h_{kl}$ and $g_{kl}$ ($l=1,\cdots, L_k$) of the vectors $\mathbf{h}_{k}$ and $\mathbf{g}_{k}$ are independent and identically distributed $\mathcal{CN}(0,1)$. Phase offsets between carrier frequencies employed by the sources and ${\rm{R}}_k$ are absorbed into $h_{kl}$ and $g_{kl}$. It is also assumed that the channels are reciprocal (this can be true when using time-division duplexing (TDD)) and static during the two transmission stages for transmitting numerous
data packets consecutively (this is true for block-fading channels). As a result, a small fraction of time used for the training symbols of CSI estimation and for feeding back the estimated CSI may essentially have little impact on the achievable transmission rate. At the beginning of the proposed scheme, one relay is selected to assist the source nodes in communication. The RS process can be implemented by using a distributed mechanism \cite{Bletsas2006}, which requires the relay node ${{\rm{R}}_k}$ to know only $\mathbf{h}_{k}$ and $\mathbf{g}_{k}$, rather than the CSI of the entire network.
Generally, the source nodes need extra feedback information from the relay to preprocess their information. This requirement can be removed under some specific configurations, which will be detailed in the sequel.

We assume that ${\rm{S}}_{1}$ and ${\rm{S}}_{2}$ share the same $M$-ary constellation alphabet set $S$, where $|S|=M$. For the source ${\rm{S}}_{i}$, $i=1,2$, the modulated symbol $s_{i}\in S$ carries $\log M$ bits which are stacked in a {\small{$1\times\log M$}} vector $W_{i}$. The symbol $s_{i}$ is transmitted with power $P_s$ in the form of {\small{$x_{i}\triangleq
\sqrt{P_s}u_{i}s_{i}$}}, where $u_{i}$ denotes the preprocessing imposed
on $s_{i}$, and we have $\left| {{u_i}} \right| = 1$. In addition, we define the mapping function $\mathcal{M}$ by which $W_{i}$ is mapped to $s_{i}$, i.e., $s_{i}=\mathcal{M}\left(W_{i}\right)$. Based on all, we define symbol-level XOR as {\small{${s_1} \odot {s_2} \triangleq  {\cal M}\left( {{{\cal M}^{ - 1}}\left( {{s_1}} \right) \oplus {{\cal M}^{ - 1}}\left( {{s_2}} \right)} \right)$}}, where $\oplus$ denotes the bitwise XOR operation, and ${\cal M}^{-1}$ is the inverse mapping of {\small{${\cal M}$}}. Then, network coded symbol (NCS) is defined as {\small{${s_{NC}} \triangleq {s_1} \odot {s_2}$}}, whose alphabet is also {\small{$S$}}.

\begin{figure}
\begin{centering}
\includegraphics[scale=0.4]{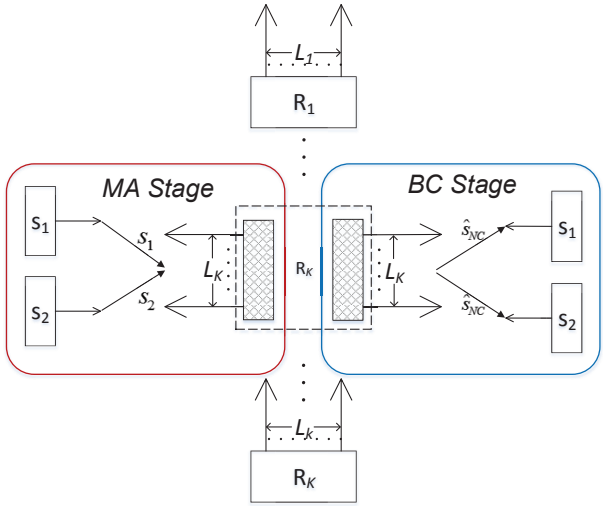}
\par\end{centering}

\caption{The TWRC system model considered: only one relay is selected and activated throughout the MA and BC stages.}

\end{figure}

\subsection{DF Based Transmission with an Arbitrary Relay ${\rm{R}}_k$}
We proceed to outline the transmission flow of the DF based TWRC. We consider the communication using an arbitrary relay $\rm{R}_{k}$ as an example. First of all, if preprocessing is employed by the sources, the relay $\rm{R}_{k}$ calculates and feeds back $u_1$ and $u_2$ to $\rm{S}_1$ and $\rm{S}_2$, respectively. The feedback is assumed to be perfect. Then, the bidirectional transmission will take place in the MA and BC stages. In the
MA stage, the signals transmitted by the sources are assumed to arrive at ${\rm{R}}_k$ simultaneously. The signal vector received at ${\rm{R}}_{{k}}$ is given
by
{\small{\begin{equation}
\mathbf{y}_{k}=\mathbf{h}_{{k}}x_{1}+\mathbf{g}_{{k}}x_{2}+\mathbf{n},\label{eq:receive signal}
\end{equation}}}where $\mathbf{n}$ is an $L_{k}\times1$ complex Gaussian noise vector obeying {\small{$\mathcal{CN}\left({\mathbf{0}_{L_k\times1},\sigma^{2}{\mathbf{I}_{L_k\times L_k}}}\right)$}}. The ML multi-user detector (MUD) is employed to jointly decode both
messages from $\mathbf{y}_{{k}}$, i.e.,
\begin{align}
&\nonumber\left({{{\hat{s}}_{1}},{{\hat{s}}_{2}}}\right)\\&=\mathop{\rm {arg}}\limits _{\left({{\widetilde{\phi}_{1}},{\widetilde{\phi}_{2}}}\right){\in}{{S}\times{S}}}{\rm{min}}\left\Vert {\mathbf{y}_{{k}}}-\sqrt{P_s}{\mathbf{h}_{{k}}}{u_{1}}\widetilde{\phi}_{1}-\sqrt{P_s}{\mathbf{g}_{{k}}}{u_{2}}\widetilde{\phi}_{2}\right\Vert ^{2}.\label{eq:multiuser detector}
\end{align}Then, ${\rm{R}}_{{k}}$ generates the NCS by using ${{\hat s}_{NC}} = {{\hat s}_1} \odot {{\hat s}_2}, \hat{s}_{NC}\in S$.
During the BC stage, ${\rm{R}}_k$ broadcasts the signal $\mathbf{X}_{R}=\sqrt{P_r}\mathbf{w}_{k}\hat{s}_{NC}$ to the sources, where $P_r$ is the broadcast power, $\mathbf{w}_{{k}}$ is an $L_{k}\times1$ beamforming vector designed to exploit the spatial diversity offered by the multiple antennas of ${\rm{R}}_k$, and $\left\Vert \mathbf{w}_{k}\right\Vert =1$.

The signals received by $\rm{S}_1$ and $\rm{S}_2$ may be expressed as
\begin{align}
\label{BCreceived}
&\nonumber y_{BC,1}=\sqrt{P_r}\mathbf{h}_{{{k}}}^{T}\mathbf{w}_{{k}}{\hat{s}}_{NC}+n_1, \\ &y_{BC,2}=\sqrt{P_r}\mathbf{g}_{{{k}}}^{T}\mathbf{w}_{{k}}{\hat{s}}_{NC}+n_2,
\end{align}respectively, where $n_1$ and $n_2$ are independent Gaussian noises following $\mathcal{CN}\left(0,\sigma^{2}\right)$. In order to estimate $\hat{s}_{NC}$, $\rm{S}_1$ and $\rm{S}_2$ employ the following ML detectors
\begin{align}
&\nonumber {{\hat{\hat{s}}_{NC,1}}}=\mathop{\rm {arg}}\limits _{{\widetilde{\phi}_{1}}{\in}{S}}\min\left|y_{BC,1}-\sqrt{P_r}{\mathbf{h}_{{k}}^{T}}\mathbf{w}_{k}\widetilde{\phi}_{1}\right|^{2} \\
&{\hat{\hat{s}}_{NC,2}}=\mathop{\rm {arg}}\limits _{{\widetilde{\phi}_{2}}{\in}{S}}\min\left|y_{BC,2}-\sqrt{P_r}{\mathbf{g}_{{k}}^{T}}\mathbf{w}_{k}\widetilde{\phi}_{2}\right|^{2},
\end{align}respectively. For $i=1,2$, the source ${\rm{S}}_{3-i}$ decides the desired information sent by ${\rm{S}}_{i}$ according to $\tilde{s}_{i}={\hat{\hat{s}}_{NC,3-i}}\odot s_{3-i}$.
\subsection{The SER Bound When Using ${\rm{R}}_k$}
For the relay ${\rm{R}}_k$, the instantaneous overall SER serves as a relevant performance metric for the RS scheme discussed later. Hence, let us analyze the instantaneous overall SER,  i.e., {\small{$\mathrm{p}_{k}^{E2E}\triangleq\frac{1}{2}\sum_{i=1}^{2}\Pr\left\{ \tilde{s}_{i}\neq s_{i}\left|\mathbf{g}_{{k}},\mathbf{h}_{{k}}\right.\right\}$}}. According to \cite{ZFSelect}, we have ${\rm {p}}_{k}^{E2E}\leq\mathrm{p}_{k}^{MA}+\frac{1}{2}\sum_{i=1}^{2}\mathrm{p}_{k,i}^{BC}$, where {\small{$\mathrm{p}_{k}^{MA}\triangleq\Pr\left\{ \hat{s}_{NC}\neq s_{NC}\left|\mathbf{g}_{{k}},\mathbf{h}_{{k}}\right.\right\} $}} and $\mathrm{p}_{k,i}^{BC}\triangleq\Pr\left\{ \hat{\hat{s}}_{NC,i}\neq\hat{s}_{NC}\left|\mathbf{g}_{{k}},\mathbf{h}_{{k}}\right.\right\} $ represent the SERs at the relay and the two sources, respectively. Furthermore, again relying on \cite{ZFSelect}, ${\rm {p}}_{k}^{E2E}$ approaches $\mathrm{p}_{k}^{MA}+\frac{1}{2}\sum_{i=1}^{2}\mathrm{p}_{k,i}^{BC}$ as the signal-to-noise ratio (SNR) increases. Thus, we use  $\mathrm{p}_{k}^{MA}+\frac{1}{2}\sum_{i=1}^{2}\mathrm{p}_{k,i}^{BC}$ to approximate ${\rm {p}}_{k}^{E2E}$ in the high-SNR scenario. In the sequel, we will analyze ${\rm {p}}_{k}^{E2E}$ by first deriving the SER bounds for $\mathrm{p}_{k}^{MA}$ and $\mathrm{p}_{k,i}^{BC}$. Then, the relevant performance metric will be extracted to enable the RS methods.

We assume that when $\left({{\widetilde{\phi}_{1}},{\widetilde{\phi}_{2}}}\right)$ is transmitted the MUD obtains the estimate $\left({{\widetilde{\phi}'_{1}},{\widetilde{\phi}'_{2}}}\right)$ of  ${{\widetilde{\phi}_{1}}\odot{\widetilde{\phi}_{2}}}\neq{{\widetilde{\phi}'_{1}}\odot{\widetilde{\phi}'_{2}}}$,  $\widetilde{\phi}_{i},\widetilde{\phi}'_{i}\in S $. Referring to ${s_{NC}} \triangleq {s_1} \odot {s_2}$, we know that the error event $\left\{ \hat{s}_{NC}\neq s_{NC}\right\}$ comprises all possible cases. Based on this observation, $\mathrm{p}^{MA}_{k}$ can be given by
{\small{\begin{equation}
{\mathrm{p}_{k}^{MA}}=\Pr\left\{ \underset{{{\widetilde{\phi}_{1}}\odot{\widetilde{\phi}_{2}}}\neq{{\widetilde{\phi}'_{1}}\odot{\widetilde{\phi}'_{2}}},\:\widetilde{\phi}_{i},\widetilde{\phi}'_{i}\in S}{\cup}\left({{\widetilde{\phi}_{1}},{\widetilde{\phi}_{2}}}\right)\rightarrow\left({{\widetilde{\phi}'_{1}},{\widetilde{\phi}'_{2}}}\right)\left|\mathbf{g}_{{k}},\mathbf{h}_{{k}}\right.\right\}.
\end{equation}}}\footnote{$\left({{\widetilde{\phi}_{1}},{\widetilde{\phi}_{2}}}\right)\rightarrow\left({{\widetilde{\phi}'_{1}},{\widetilde{\phi}'_{2}}}\right)$ denotes the event that when $\left({{\widetilde{\phi}_{1}},{\widetilde{\phi}_{2}}}\right)$ is transmitted, $\left({{\widetilde{\phi}'_{1}},{\widetilde{\phi}'_{2}}}\right)$ is obtained as its estimate.}Then, the union bound is used to approximate $\mathrm{p}^{MA}_{k}$ as
{\small{\begin{align}
\label{CP_SER}
&\nonumber{\mathrm{p}_{k}^{MA}}\leq \underset{{{\widetilde{\phi}_{1}}\odot{\widetilde{\phi}_{2}}}\neq{{\widetilde{\phi}'_{1}}\odot{\widetilde{\phi}'_{2}}},\:\widetilde{\phi}_{i},\widetilde{\phi}'_{i}\in S}{\sum}\frac{1}{M^{2}}\\&\hspace{20pt}\cdot Q\left(\sqrt{\frac{\mu}{2}}\|\mathbf{g}_{{k}}u_{2}\left(\widetilde{\phi}_{2}-\widetilde{\phi}'_{2}\right)+\mathbf{h}_{{k}}u_{1}\left(\widetilde{\phi}_{1}-\widetilde{\phi}'_{1}\right)\|\right)\\
	&=\underset{\left(d_{1},d_{2}\right)\in D}{\sum}\frac{1}{M^{2}}Q\left(\sqrt{\frac{\mu}{2}}\lambda\left(u_1d_{1},u_2d_{2}\right)\right)\leq\frac{\left|D\right|}{M^{2}}Q\left(\sqrt{\frac{\mu}{2}}\lambda_{LB}\right),
\end{align}}}where $\mu\triangleq\frac{P_s}{\sigma^2}$ is the SNR, {\small{$D\triangleq\left\{ \left(d_{1},d_{2}\right)\left|d_{i}=\widetilde{\phi}_{i}-\widetilde{\phi}'_{i},{{\widetilde{\phi}_{1}}\odot{\widetilde{\phi}_{2}}}\neq{{\widetilde{\phi}'_{1}}\odot{\widetilde{\phi}'_{2}}},\:\widetilde{\phi}_{i},\widetilde{\phi}'_{i}\in S\right.\right\}
 $}}, $\lambda\left(u_1d_1,u_2d_2\right)\triangleq\|\mathbf{g}_{{{k}}}u_{2}d_{1}+\mathbf{h}_{{{k}}}u_{1}d_{2}\|$ is the DD in the MA stage, and $\lambda_{LB}$ denotes the lower bound of $\lambda\left(u_1d_{1},u_2d_{2}\right)$,$\left(d_{1},d_{2}\right)\in D$. The first inequality given in (\ref{CP_SER}) is based on the union bound whose tightness is well confirmed by \cite{tse2005fundamentals}, while the tightness of the second upper bound in (\ref{CP_SER}) depends on the design of $\lambda_{LB}$, which will be given later.

On the other hand, the SER expressions in the BC stage conditioned on the instantaneous channel state are given by
\begin{align}
\label{PBC}
\nonumber &\mathrm{p}_{k,1}^{BC}=C_{1}Q\left(\sqrt{C_{2}p\mu\left|\mathbf{w}_{{k}}^{T}\mathbf{h}_{{k}}\right|^{2}}\right),\\&\mathrm{p}_{k,2}^{BC}=C_{1}Q\left(\sqrt{C_{2}p\mu\left|\mathbf{w}_{{k}}^{T}\mathbf{g}_{{k}}\right|^{2}}\right),
\end{align}where $p\triangleq\frac{P_{r}}{P_{s}}$, $C_{1}$ and $C_{2}$ are constants depending on the modulation. Generally, $\sqrt{C_{2}\left|\mathbf{w}_{{k}}^{T}\mathbf{h}_{{k}}\right|^{2}}$ and $\sqrt{C_{2}\left|\mathbf{w}_{{k}}^{T}\mathbf{g}_{{k}}\right|^{2}}$ are DDs in BC stage.

From (\ref{CP_SER}) and (\ref{PBC}),   ${\rm{p}}^{E2E}_{k}\leq\mathrm{p}^{MA}_{k}+\frac{1}{2}\sum_{i=1}^{2}\mathrm{p}^{BC}_{k,i}$ is bounded \cite{ZFSelect} by
{\small{\begin{equation}
\label{P_bound}
{\rm {p}}_{k}^{E2E}\leq\alpha Q\left(\sqrt{\mu\min\left\{ \frac{\lambda_{LB}^{2}}{2},\, C_{2}p\left|\mathbf{w}_{k}^{T}\mathbf{h}_{k}\right|^{2},\, C_{2}p\left|\mathbf{w}_{k}^{T}\mathbf{g}_{k}\right|^{2}\right\} }\right),
\end{equation}}}where $\alpha=\frac{\left|D\right|}{M^{2}}+C_{1}
 $ is a modulation-specific constant. Since the accuracy of (\ref{PBC}) is ensured by the existing performance analysis \cite{tse2005fundamentals},  its substitution into ${\rm{p}}^{E2E}_{k}\leq\mathrm{p}^{MA}_{k}+\frac{1}{2}\sum_{i=1}^{2}\mathrm{p}^{BC}_{k,i}$ does not impact the tightness of ${\rm{p}}^{E2E}_{k}$. Hence, the tightness of the upper bound given in (\ref{P_bound}) depends only on the design of $\lambda_{LB}$. We will propose a preprocessing scheme in Section III in order to find a desired $\lambda_{LB}$. Upon using the obtained $\lambda_{LB}$, the upper bound given by the right-hand side of (\ref{P_bound}) exhibits a reasonably small gap\footnote{Due to page limitations, numerical results illustrating this gap are not included in this paper.}, especially in high-SNR region, from the actual ${\rm {p}}_{k}^{E2E}$. As a beneficial result,  the upper bound of ${\rm {p}}_{k}^{E2E}$ in (\ref{P_bound}) may be used as a relay selection metric for achieving full diversity gain.

More specifically,  we propose to select the $\hat{k}$th relay by minimizing the upper bound of ${\rm {p}}_{k}^{E2E}$ which is given by (\ref{P_bound}). Then, we have
{\small{\begin{equation}
\label{P_maxmin}
\hat{k}=\arg\underset{k=1,\ldots,K}{\max}\min\left\{ \frac{\lambda_{LB}^{2}}{2},\, C_{2}p\left|\mathbf{w}_{k}^{T}\mathbf{h}_{k}\right|^{2},\, C_{2}p\left|\mathbf{w}_{k}^{T}\mathbf{g}_{k}\right|^{2}\right\}.
\end{equation}}}This RS approach is not optimal. Because in each instant the criterion (\ref{P_maxmin})
optimizes the upper bound of the SER performance, rather than the actual SER performance. The goal of this paper is to find a simple suboptimal RS approach capable of achieving full diversity gain. In the next section, we will design $u_1$, $u_2$ and $\mathbf{w}_{k}$ in order to ensure that the tightness of the upper bound given in (\ref{P_bound}) does not impose a negative impact on the achievable diversity performance. The designs on $u_1$, $u_2$ and $\mathbf{w}_{k}$ are based on  the PR strategy. As a result, the RS approach specified by (\ref{P_maxmin}) is actually the MaxMin-RS with respect to channel gains. The PR strategy conceived not only assists the MaxMin-RS in achieving full diversity, but also allows us to obtain explicit analytical results. Notably, our analysis confirms that the proposed scheme obtains full diversity gain, and achieves better array power gain than the existing schemes.

\section{The PR AIDED MAXMIN-RS Scheme}
In this section, we first explain how the PR strategy shapes $\lambda\left(u_{1}d_{1},u_{1}d_{2}\right)$, $\left|\mathbf{w}_{k}^{T}\mathbf{h}_{k}\right|^{2}$ and $\left|\mathbf{w}_{k}^{T}\mathbf{g}_{k}\right|^{2}$ in the PR preprocessing stage for the MA stage, and the PR beamforming stage for the BC stage, respectively. The two operations provide the desired lower bounds for $\lambda\left(u_{1}d_{1},u_{2}d_{2}\right)$ and $\left|\mathbf{w}_{k}^{T}\mathbf{h}_{k}\right|^{2}$ (and $\left|\mathbf{w}_{k}^{T}\mathbf{g}_{k}\right|^{2}$), respectively. These desired lower bounds reshape the upper bound of the instantaneous end-to-end SER ${\rm {p}}_{k}^{E2E}$ in (\ref{P_bound}), which is optimized relaying on by the RS approach of (\ref{P_maxmin}) to achieve full diversity order.
As a result, the actual averaged end-to-end performance is also confirmed to exhibit full diversity order.

\subsection{PR strategy}
\subsubsection{PR Preprocessing in the MA stage}Assume that ${\rm{R}}_k$ is selected. Considering the DDs $\lambda\left(u_1d_1,u_2d_2\right)$, $\left(d_{1},d_{2}\right)\in D$, which determines ${\mathrm{p}_{k}^{MA}}$ of (\ref{CP_SER}),
we propose the PR preprocessing to choose $u_1$ and $u_2$ such that a reasonable lower bound on $\lambda\left(u_1d_1,u_2d_2\right)$ is given. More specifically, the motivation of the PR preprocessing comes from the observation that without any preprocessing, when $d_{1}\neq0, d_{2}\neq0$, $\lambda\left(u_1d_1,u_2d_2\right)$ can be lower bounded by
\begin{align}
\label{withoutPR}
&\nonumber\lambda^{2}\left(u_1d_{1},u_2d_{2}\right)=\left\Vert \mathbf{g}_{{k}}d_{2}\right\Vert ^{2}+\left\Vert \mathbf{h}_{{k}}d_{1}\right\Vert ^{2}+2\Re\left\{ \mathbf{h}_{{k}}^{H}\mathbf{g}_{{k}}d_{1}^{\text{*}}d_{2}u_{1}^{\text{*}}u_{2}\right\} \\ \nonumber
	&=\left\Vert \mathbf{g}_{{k}}d_{2}\right\Vert ^{2}+\left\Vert \mathbf{h}_{{k}}d_{1}\right\Vert ^{2}+2\left|\mathbf{h}_{{k}}^{H}\mathbf{g}_{{k}}d_{1}^{\text{*}}d_{2}\right|\cos\left(\angle d_{2}-\angle d_{1}+\varphi_{k}\right)\\ \nonumber
	&\geqslant2\left\Vert \mathbf{g}_{{k}}d_{2}\right\Vert \left\Vert \mathbf{h}_{{k}}d_{1}\right\Vert -2\left\Vert \mathbf{g}_{{k}}d_{2}\right\Vert \left\Vert \mathbf{h}_{{k}}d_{1}\right\Vert \left|\cos\left(\angle d_{2}-\angle d_{1}+\varphi_{k}\right)\right|\\
	&\geqslant2\Gamma_{k}d_{min}^{2}\left(1-\left|\cos\left(\angle d_{2}-\angle d_{1}+\varphi_{k}\right)\right|\right),
 \end{align}where $\varphi_k\triangleq\angle\frac{\mathbf{h}_{{{k}}}^{H}\mathbf{g}_{{{k}}}}{|\mathbf{h}_{{{k}}}^{H}\mathbf{g}_{{{k}}}|}
 $, referred to as the EA of ${\rm{R}}_{k}$, is  the angle between $\mathbf{h}_{{{k}}}$ and $\mathbf{g}_{{{k}}}$, $\Gamma_{k}\triangleq\min\left\{ ||\mathbf{h}_{k}||^{2},||\mathbf{g}_{k}||^{2}\right\}
 $, and $d_{min}$ is the minimum value of all possible nonzero values of  $\left|d_1\right|$ and $\left|d_2\right|$. Note that $\varphi_k$ is uniformly distributed over $\left[0,2\pi\right]$.
This makes $\angle d_{2}-\angle d_{1}+\varphi_{k}$ also uniformly distributed. Hence, there is a nonnegligible probability that the lower bound in (\ref{withoutPR}) becomes close to 0. This event then dominates ${\mathrm{p}_{k}^{MA}}$ is given by (\ref{CP_SER}).

In order to decrease ${\mathrm{p}_{k}^{MA}}$, we eliminate the randomness of $\varphi_k$ by judiciously designing $u_1$ and $u_2$ as
\begin{equation}
\label{PR-MA}
u_{1}=\exp\left(j\left(\upsilon+\varphi_{k}\right)\right),\, u_{2}=1,
\end{equation}where $j=\sqrt{-1}$, and $\upsilon$ is a constant rotation angle that is specified later. For the sake of explicit clarity, the geometrical interpretation of this PR preprocessing operation is illustrated in Fig. 2. Furthermore, the effect of PR is analytically confirmed by the following proposition:
\begin{figure}
\begin{centering}
\includegraphics[scale=0.72]{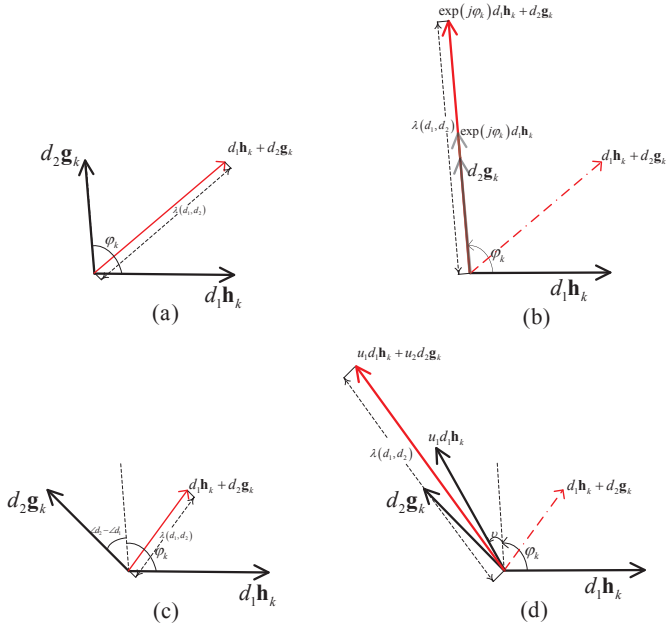}
\par\end{centering}

\caption{Subfigures (a) and (b) describe the PR preprocessing operation in the special case of $\angle d_{1}=\angle d_{2}=0$. Subfigure (a) illustrates $\lambda\left(u_1d_1,u_2d_2\right)$ without any preprocessing in this special case, highlighting how $\varphi_k$ impacts  $\lambda\left(u_1d_1,u_2d_2\right)$. It is observed from subfigure (b) that once $\varphi_k$ is pre-canceled, $\exp\left(j\varphi_{k}\right)\mathbf{h}_{k}d_1$ and $\mathbf{g}_{k}d_2$ coincide along the same direction, hence $\lambda\left(u_{1}d_1,u_{2}d_2\right)$ is improved. Generally, there always exists $\left(d_{1},d_{2}\right)\in D$ with $\angle d_{2}-\angle d_{1}\neq0$, thus both $\varphi_k$ and $\angle d_{2}-\angle d_{1}$ are involved in determining $\lambda\left(u_{1}d_1,u_{2}d_2\right)$ as shown in
 subfigure (c). Subfigure (d) illustrates how $\lambda\left(u_{1}d_1,u_{2}d_2\right)$ is increased by the PR preprocessing in this general case. Subfigure (d) shows the rotation of two angles $\varphi_k$ and $\upsilon$, while the rotation of $\varphi_k$ aims to cope with the angle between $\mathbf{h}_{k}$ and $\mathbf{g}_{k}$.The rotation of $\upsilon$ on the other hand intends to adjust the influence of $\angle d_{2}-\angle d_{1}$ on $\lambda\left(u_{1}d_1,u_{2}d_2\right)$. Finally, comparing subfigures (c) and (d), we see that $\lambda\left(u_{1}d_1,u_{2}d_2\right)$ is increased by the PR preprocessing operation.}

\end{figure}

\begin{prop}
With the PR preprocessing, we have $\lambda_{LB}=\sqrt{C_{3}\Gamma_{k}\left|d_{min}\right|^{2}}$, i.e., $\lambda\left(u_{1}d_1,u_{2}d_2\right)\geqslant\sqrt{C_{3}\Gamma_{k}\left|d_{min}\right|^{2}}$ for all $\left(d_{1},d_{2}\right)\in D$. As a result, ${\mathrm{p}_{k}^{MA}}$ is upper bounded as
\begin{equation}
\label{lowerbound1}
\mathrm{p}_{k}^{MA}\leqslant C_{4}Q\left(\sqrt{\frac{\mu}{2}C_{3}\Gamma_{k}\left|d_{min}\right|^{2}}\right),
\end{equation}where {\small{$C_{3}=\min\left(1,\rho_{\min}\left(\upsilon\right)\right)$ with $\rho_{\min}\left(\upsilon\right)\triangleq\underset{\left(d_{1},d_{2}\right)\in D}{\min}2\left(1-\left|\cos\left(\angle d_{1}-\angle d_{2}+\upsilon\right)\right|\right)$}}, and $C_4=\frac{\left|D\right|}{M^{2}}$.
\end{prop}
\begin{IEEEproof}
Please see Appendix A.
\end{IEEEproof}
As shown in Proposition 1, $C_{3}=\min\left(1,\rho_{\min}\left(\upsilon\right)\right)$ depends only on $\upsilon$. Hence, $\upsilon$ controls the upper bound of $\mathrm{p}_{k}^{MA}$. If  $\rho_{\min}\left(\upsilon\right)$ approaches $0$, $C_{3}=\min\left(1,\rho_{\min}\left(\upsilon\right)\right)$ will also approach $0$, which may trivialize the upper bound of $\mathrm{p}_{k}^{MA}$ given in (\ref{lowerbound1}). Therefore, we constrain $\upsilon$ so that $\rho_{\min}\left(\upsilon\right)>0$ to avoid the occurrence of  $C_3=0$. Note that $\upsilon$ can be easily obtained in an off-line manner\footnote{The optimal value of $\upsilon$ can be calculated off-line by
$\upsilon=\arg\mathop{\max}\limits _{\upsilon'\in\left[{0,2\pi}\right]}\rho_{\min}\left(\upsilon'\right).$ } .

\subsubsection{PR Aided Bidirectional Beamforming in the BC stage}For the BC stage, we would also like to make $\min\left\{ \left|\mathbf{w}_{k}^{T}\mathbf{h}_{k}\right|^{2},\left|\mathbf{w}_{k}^{T}\mathbf{g}_{k}\right|^{2}\right\}$ be determined by $\Gamma_k$. This requirement allows $\min\left\{ \left|\mathbf{w}_{k}^{T}\mathbf{h}_{k}\right|^{2},\left|\mathbf{w}_{k}^{T}\mathbf{g}_{k}\right|^{2}\right\}$ and $\frac{\lambda_{LB}^{2}}{2}$ to be simultaneously optimized by a RS approach regarding $\Gamma_k$. Correspondingly, both ${\rm{p}}^{MA}_{k}$ in (\ref{CP_SER}) and ${\rm{p}}^{BC}_{k,i}$ in (\ref{PBC}) can be improved by the RS approach. According to this requirement, we propose a broadcast scheme detailed as follows:

 If $L_{{k}}=1$, there is no need to design the beamforming vector $\mathbf{w}_{{k}}$. According to (\ref{PBC}), for this case, we have $\min\left\{ \left|\mathbf{w}_{k}^{T}\mathbf{h}_{k}\right|^{2},\left|\mathbf{w}_{k}^{T}\mathbf{g}_{k}\right|^{2}\right\}=\Gamma_{k}$ and
\begin{equation}
\label{BClowerbound}
\mathrm{p}_{k,i}^{BC}\leq C_{1}Q\left(\sqrt{C_{2}p\mu\Gamma_{k}}\right).
\end{equation}For the case of $L_{{k}}\geqslant2$, the PR beamforming weights can be obtained as
\begin{equation}
\label{transmitted vector}
\mathbf{w}_{{k}}=\sqrt{\frac{1}{2}}\left(\exp\left(-j\varphi_{k}\right)\mathbf{q}_{1}^{H}+j\mathbf{q}_{2}^{H}\right),
\end{equation}where $\varphi_k=\angle\frac{\mathbf{h}_{{{k}}}^{H}\mathbf{g}_{{{k}}}}{|\mathbf{h}_{{{k}}}^{H}\mathbf{g}_{{{k}}}|}$, $\mathbf{q}_{1}$ and $\mathbf{q}_{2}$ result from the Gram-Schmidt orthogonalization that is given by
\begin{equation}
\label{GS}
\mathbf{q}_{1}^{T}=\frac{1}{r_{11}}\mathbf{h}_{{k}},\:\mathbf{q}_{2}^{T}=\frac{\mathbf{g}_{k}-r_{12}\mathbf{q}_{1}^{T}}{r_{22}},
\end{equation}in which we have $r_{11}=\left\Vert \mathbf{h}_{k}\right\Vert$,  $r_{12}=\frac{\mathbf{h}_{k}^{H}\mathbf{g}_{k}}{\left\Vert \mathbf{h}_{k}\right\Vert }$ and $r_{22}=\left\Vert \mathbf{g}_{k}-\frac{\mathbf{h}_{k}^{H}\mathbf{g}_{k}}{\left\Vert \mathbf{h}_{k}\right\Vert ^{2}}\mathbf{h}_{{k}}\right\Vert $. The effect of this choice of $\mathbf{w}_{{k}}$ is given as follows.
\begin{prop}
For the case of $L_{k}\geqslant2$, when $\mathbf{w}_{{k}}$ is chosen according to (\ref{transmitted vector}), we have {\small{$\min\left\{ \left|\mathbf{w}_{k}^{T}\mathbf{h}_{k}\right|^{2},\left|\mathbf{w}_{k}^{T}\mathbf{g}_{k}\right|^{2}\right\} =\frac{1}{2}\Gamma_{k}
 $}}. Then, a general upper bound of  $\mathrm{p}_{k,i}^{BC}$ is given by
\begin{equation}
\label{bcl}
\mathrm{p}_{k,i}^{BC}\leq C_{1}Q\left(\sqrt{\frac{C_{2}p\mu}{2}\Gamma_{k}}\right).
\end{equation}
\end{prop}
\begin{IEEEproof}
Please see Appendix B. \end{IEEEproof} Jointly considering (\ref{bcl}) and (\ref{BClowerbound}), we can see that a RS approach relying on $\Gamma_{k}$ is capable of lowering the upper bound of $\mathrm{p}_{k,i}^{BC}$. Based on this insight, an RS approach is given as follows.
\subsection{Relay Selection}
Propositions 1 and 2 state that the upper bounds of ${\rm{p}}^{BC}_{k,i}$ and ${\rm{p}}^{MA}_k$ are determined by $\Gamma_{k}$ with some modulation-specific constants. This insight indicates that using $\Gamma_{k}$ as the selection weight throughout both the MA and BC stages reduces ${\rm{p}}^{BC}_{k,i}$ and ${\rm{p}}^{MA}_k$, which in turn improves the overall end-to-end performance. Based on this insight, the RS approach in (\ref{P_maxmin})  becomes
\begin{equation}
\hat{k}=\arg\underset{_{k=1,\ldots,K}}{\max}\Gamma_{k}.\label{eq:RSselection}
\end{equation}Since $\Gamma_{k}=\min\left\{||\mathbf{h}_{k}||^{2},||\mathbf{g}_{k}||^{2}\right\}
 $, the proposed criterion in (\ref{eq:RSselection}) constitutes a MaxMin optimization problem. Interestingly, we note that the existing MaxMin RS criterion used in single-antenna relay networks \cite{jing2009relay,AF-SER,zhou2010decode,eslamifarmax,caospnc,krikidis2010relay} becomes a special case of the proposed general MaxMin criterion.

\section{Diversity and SER Analysis of the PR-MaxMin-RS Scheme}

Achieving full diversity in the end-to-end SER is the main design goal of this paper. In particular, the end-to-end diversity order is defined as
$d\triangleq\underset{\mu \rightarrow\infty}{\lim}-\frac{\log P_{E}}{\log \mu}$,
where $P_{E}$ denotes the average overall end-to-end SER, and it is generally limited by the worst among
$P_{MA}$, $P_{BC,1}$, and $P_{BC,2}$. These three constituent error probabilities denote the average SER achieved by the relay in the MA stage, and by the sources $\rm{S}_1$ and $\rm{S}_2$ in the BC stage, respectively. In this section, we first elaborate on how the proposed scheme achieves full diversity gain. Then, we provides an explicit SER analysis in order to further confirm the advantage of the proposed scheme. The given analytical results are applicable to the general antenna configuration. Additionally, we also discuss two special cases which exhibit some some interesting properties.
\subsection{Diversity Analysis}
In order to present the diversity performance, we first investigate the upper bound of the end-to-end SER ${\rm{p}}^{E2E}_k$.
Based on (\ref{bcl}) and (\ref{lowerbound1}), {\small{${\rm{p}}^{E2E}_k=\frac{1}{2}\sum_{i=1}^{2}\left(\mathrm{p}^{MA}_{k}+\mathrm{p}^{BC}_{k,i}\right)$}} is bounded as
\begin{equation}
\label{e2e}
\mathrm{p}_{k}^{E2E}\leq\alpha Q\left(\sqrt{\beta\mu\Gamma_{k}}\right),
\end{equation}where $\beta$ is a constant depending on $C_2$, $C_3$ and $C_4$. The averaged SER is
\begin{align}
\label{e2eb}
P_{E}&=\mathbb{E}\left\{ \mathrm{p}_{\hat{k}}^{E2E}\right\} \leq\alpha\underset{\Gamma_{\hat{k}}}{\mathbb{E}}\left\{Q\left(\sqrt{\beta\mu\Gamma_{\hat{k}}}\right)\right\}\leq\alpha\underset{\xi}{\mathbb{E}}\left\{Q\left(\sqrt{\beta\mu\xi}\right)\right\}\notag\\&\propto\frac{1}{\mu^{L}},
\end{align}
where the second inequality of (\ref{e2eb}) holds owing to the fact that
{\small{$\Gamma_{\hat{k}}\geqslant\xi\triangleq\underset{_{k=1,\ldots,K}}{\max}\underset{_{l=1,\ldots,L_{k}}}{\max}\min\left\{ |h_{kl}|^{2},|g_{kl}|^{2}\right\}$}} (see the proof in Appendix C) and the property of $\xi$ that $\alpha\underset{\xi}{\mathbb{E}}\left\{Q\left(\sqrt{\beta\mu\xi}\right)\right\}\propto\frac{1}{\mu^{L}}$ (see the proof in Appendix C). The right side of (\ref{e2eb}) that implies $d\triangleq-\frac{\log P_{E}}{\log\mu}\geqslant L$. Due to the $\textit{a priori}$ knowledge that $d\leqslant L$, the result $d=L$ is confirmed.

\emph{Remark}: We have proved that with the aid of the PR strategy is able to achieve full diversity gain. Furthermore,
we will also show in the sequel that without the PR strategy, the full diversity gain cannot be achieved by only applying
the MaxMin criterion straightforwardly to the general configuration ($K\geqslant2$, non-binary modulation). The performance enhancement obtained in the former scheme comes from the fact that the PR preprocessing and PR beamforming operations
guarantee that the full diversity gain is achieved in the MA and BC stages, respectively.

We also observe that for some special configurations, even if the PR preprocessing is not invoked, full diversity gain can still be achieved in the MA stage. Hence, the PR preprocessing operation and the corresponding overhead spent for feeding back $u_1$ can be removed in such scenario. More details can be found in the following two propositions.

\begin{prop}
For the scenario with a single multi-antenna relay, i.e., $K=1$, and arbitrary modulations, there is no need to preprocess the transmitted signal of sources for obtaining the full diversity gain. In other words, we may have $u_{1}=1, u_{2}=1$, and the full diversity order $d=L$ can still be guaranteed.
\end{prop}
\begin{IEEEproof}
In this scenario, all of $L$ antennas are gathered in one relay. The transmission in the MA stage is equivalent to being happened in a $2\times L$ MIMO system, where it is well known that the ML-based MUD in (\ref{eq:multiuser detector}) can obtain full diversity without any preprocessing \cite{tse2005fundamentals}
\end{IEEEproof}
\begin{prop}
For the scenario employing BPSK modulation and arbitrary antenna configurations, there is no need to preprocess the transmitted signal of sources for obtaining the full diversity gain. In other words, we may have $u_{1}=1, u_{2}=1$, the full diversity order $d=L$ can still be ensured.
\end{prop}
\begin{IEEEproof}
For BPSK modulation, by enumerating all possible cases in $D$, it is readily observed that there is
no element of $\left\{ d_{1}\neq0,\, d_2\neq0\right\}$ belonging to $D$. In other words, $d_{1}d_{2}=0$ always holds true, which results in $\lambda^{2}\left(u_1d_1,u_2d_2\right)=||\mathbf{g}_{{\hat{k}}}d_{2}||^{2}+||\mathbf{h}_{{\hat{k}}}d_{1}||^2$. Then,
$\lambda\left(u_1d_1,u_2d_2\right)\geqslant\sqrt{\Gamma_{k}\left|d_{min}\right|^{2}}$ holds in spite of any preprocessing. Following the similar operation on (\ref{e2eb}), we see that $d=L$ for BPSK modulation even without any preprocessing.
\end{IEEEproof}

\subsection{SER Analysis}
In order to further confirm the proposed scheme's advantage in terms of array power gain, we present the SER analysis in this subsection. For simplicity, we focus on the scenario where all $K$ relays are equipped with the same number of antennas, i.e, $L_{1}=L_{2}=\cdots L_{K}=\mathcal{L}$. In order to get $P_E$, we firstly note that $P_{E}\leq P_{MA}+\frac{1}{2}\sum_{i=1}^{2}P_{BC,i}$ \cite{ZFSelect}. Additionally, we assume that  $\mathbf{g}_{{\hat{k}}}$ and $\mathbf{h}_{{\hat{k}}}$ follow the same distribution, hence $P_{BC,1}=P_{BC,2}$. Therefore, we obtain
\begin{equation}
\label{P_E}
P_{E}\leq P_{MA}+P_{BC,1}.
\end{equation}It is noted that $ P_{MA}$ depends on the particular modulation and the corresponding choice of $\upsilon$ in the PR preprocessing scheme. Hence, considering the MPAM modulation as an example, we provide the following proposition about the choice of $\upsilon$ and the corresponding $P_E$.
\begin{prop}
For MPAM modulation and the PR preprocessing operation characterized by (\ref{PR-MA}), $\upsilon=\frac{\pi}{2}$ is the best choice for maximizing $\lambda_{LB}$. With this particular choice of $\upsilon$,  $P_{MA}$ and $P_{BC,1}$ are given as follows:
{\small{\begin{align}
\label{PMA_F}
\nonumber&{P_{MA}}\leq\\&\underset{\left(d_{1},d_{2}\right)\in D\left(s_{NC},\hat{s}_{NC}\right)}{\sum}\frac{1}{M^2\pi}\int_{0}^{\frac{\pi}{2}}\psi_{\gamma_{1}}\left(\frac{\left\Vert d_{1}\right\Vert ^{2}}{4\sin^{2}\theta}\right)\psi_{\gamma_{1}}\left(\frac{\left\Vert d_{2}\right\Vert ^{2}}{4\sin^{2}\theta}\right)d\theta,
 \end{align}}}
{\small{\begin{align}
\label{bc_pam1}
\nonumber &P_{BC,1}=\\&\begin{cases}
\begin{array}{cc}
\frac{2\left(M-1\right)}{M\pi}\int_{0}^{\frac{\pi}{2}}\psi_{\gamma_{1}}\left(\frac{3p}{\sin^{2}\theta\left(M^{2}-1\right)}\right)d\theta, & \mathcal{L}=1\\
\frac{2\left(M-1\right)}{M\pi}\int_{0}^{\frac{\pi}{2}}\psi_{\gamma_{1}}\left(\frac{3p}{2\sin^{2}\theta\left(M^{2}-1\right)}\right)d\theta, &\mathcal{L}>1
\end{array}\end{cases}.
\end{align}}}where $\psi_{\gamma_{1}}\left(t\right)$ is the moment generating function (MGF) of $\gamma_{1}=\mu\left\Vert \mathbf{h}_{{\hat{k}}}\right\Vert ^{2}$, and the definition of $\psi_{\gamma_{1}}\left(t\right)$ is $\psi_{\gamma_{1}}\left(t\right)=\int_{0}^{\infty}f_{\gamma_{1}}\left(\gamma\right)\exp\left(t\gamma\right)d\gamma$, in which $f_{\gamma_{1}}\left(\gamma\right)$ is the pdf of variable $\gamma_{1}$, and it is given as (\ref{mgf}).
\begin{figure*}[t]
{\small{\begin{align}
\label{mgf}
\nonumber \psi_{\gamma_{1}}\left(t\right)&=K\sum_{k=0}^{K-1}\left(\begin{array}{c}
K-1\\
k
\end{array}\right)\left(-1\right)^{k}\left(\mathcal{F}\left(t,2k\right)\right)-2K\left(K-1\right)\sum_{k=0}^{K-2}\left(\begin{array}{c}
K-2\\
k
\end{array}\right)\left(-1\right)^{k}\left(\mathfrak{L}\left(t,2k+1\right)-\mathfrak{L}\left(t,2k+2\right)\right)
\end{align}}}where
{\small{\begin{align}
\nonumber &\mathcal{F}\left(t,k\right)=\sum_{\begin{array}{c}
j_{0},j_{2},\ldots j_{\mathcal{L}-1}\\
j_{0}+j_{2}+\ldots j_{\mathcal{L}-1}=k
\end{array}}^{k}\frac{\left(k\right)!\prod_{l=0}^{\mathcal{L}-1}\left(\frac{1}{l!}\right)^{j_{l}}}{j_{0}!j_{2}!\ldots j_{\mathcal{L}-1}!}\frac{\left(\mathcal{L}-1+\sum_{l=0}^{\mathcal{L}-1}lj_{l}\right)!}{\left(\mathcal{L}-1\right)!}\left(\frac{1}{k+1+t\mu}\right)^{\mathcal{L}+\sum_{l=0}^{\mathcal{L}-1}lj_{l}}\\
\nonumber &\mathfrak{L}\left(t,k\right)=\sum_{\begin{array}{c}
j_{0},j_{2},\ldots j_{\mathcal{L}-1}\\
j_{0}+j_{2}+\ldots j_{\mathcal{L}-1}=k
\end{array}}^{k}\frac{\left(k\right)!\prod_{l=0}^{\mathcal{L}-1}\left(\frac{1}{l!}\right)^{j_{l}}}{j_{0}!j_{2}!\ldots j_{\mathcal{L}-1}!}\\ &\frac{\left(\frac{1}{k+1}\right)^{\mathcal{L}+\sum_{l=0}^{\mathcal{L}-1}lj_{l}}}{\left(\left(\mathcal{L}-1\right)!\right)^{2}}\left(\mathcal{L}-1+\sum_{l=0}^{\mathcal{L}-1}lj_{l}\right)!\left(\frac{\left(\mathcal{L}-1\right)!}{\left(1+t\mu\right)^{\mathcal{L}}}-\sum_{l=0}^{\mathcal{L}-1+\sum_{l=0}^{\mathcal{L}-1}lj_{l}}\left(\frac{1}{2+t\mu}\right)^{l+\mathcal{L}}\frac{\left(\mathcal{L}+l-1\right)!}{l!}\right)
\end{align}}}
\rule{18cm}{0.1em}
\end{figure*}
\end{prop}
\begin{IEEEproof}
Please see Appendix D.
\end{IEEEproof}Summing up (\ref{bc_pam1}) and (\ref{PMA_F}), an upper bound of $P_E$ is given by (\ref{P_E}), whose accuracy will be further confirmed by the numerical results presented later. From the proof of Proposition 5, it is noted that the SNR $\mu$ is always weighted by $\left\Vert \mathbf{h}_{{\hat{k}}}\right\Vert ^{2}$ or $\left\Vert \mathbf{g}_{{\hat{k}}}\right\Vert ^{2}$ in the instantaneous SER bound. Taking $\left\Vert \mathbf{h}_{{\hat{k}}}\right\Vert ^{2}$ for example, we have $\left\Vert \mathbf{h}_{{\hat{k}}}\right\Vert ^{2}=\sum_{l=1}^{\mathcal{L}}\left|h_{{\hat{k}},l}\right|^{2}$, where $h_{{\hat{k}},1},h_{{\hat{k}},2},\ldots,h_{{\hat{k}},\mathcal{L}}$ independently follow the identical distribution. Hence, the fading of $\mathcal{L}$ channels characterised by  $h_{{\hat{k}},1},h_{{\hat{k}},2},\ldots,h_{{\hat{k}},\mathcal{L}}$ respectively are averaged by $\left\Vert \mathbf{h}_{{\hat{k}}}\right\Vert ^{2}$. This insight can be applied to $\left\Vert \mathbf{g}_{{\hat{k}}}\right\Vert ^{2}$ similarly. Therefore, an improved array power gain can be observed from the SER performance achieved because $\left\Vert \mathbf{h}_{{\hat{k}}}\right\Vert ^{2}$ or $\left\Vert \mathbf{g}_{{\hat{k}}}\right\Vert ^{2}$ collects the path diversity of $\mathcal{L}$ channels. In addition, since the stochastic distribution of $\left\Vert \mathbf{h}_{{\hat{k}}}\right\Vert ^{2}$ has been given in Proposition 5, the SER result of Proposition 5 may be readily extended to other modulations schemes.

\subsection{Array Power Gain Analysis Based on an Upper Bound of $P_{E}$}

The above SER analysis implies that the proposed scheme enjoys advantages in the achievable array power gain. In this subsection, we provide an asymptotic analysis in order to understand the achievable array power gain inherent in the proposed scheme. Since the stochastic property of the selected channel is complex, it is difficult to obtain asymptotic result
of the actual average SER. Alternatively, we opt for relaxing the actual average SER to a desired upper bound of the average SER, whose physical significance is very clear. From the asymptotic analysis based on the upper bound, we obtain further insights into the array power gain and diversity order of the proposed scheme.
For the sake of simple description of the upper bound, we first define $\widetilde{\Gamma}_{k}=\sum_{l=1}^{L_{k}}\min\left\{ |h_{kl}|^{2},|g_{kl}|^{2}\right\}$ and $\widetilde{k}=\arg\underset{_{k=1,\ldots,K}}{\max}\widetilde{\Gamma}_{k}$. Then, we have
\begin{equation}\label{AW_P}
\Gamma_{\hat{k}}\geq\widetilde{\Gamma}_{\widetilde{k}},
\end{equation}which is obtained by invoking
\begin{align}
\Gamma_{k}&=\min\left\{ ||\mathbf{h}_{k}||^{2},||\mathbf{g}_{k}||^{2}\right\} =\min\left\{ \sum_{l=1}^{L_{k}}|h_{kl}|^{2},\sum_{l=1}^{L_{k}}|g_{kl}|^{2}\right\} \notag\\&\geqslant\sum_{l=1}^{L_{k}}\min\left\{ |h_{kl}|^{2},|g_{kl}|^{2}\right\} =\widetilde{\Gamma}_{k}.
\end{align}
From (\ref{AW_P}), the upper bound of the average SER is given by
\begin{equation}\label{A_U_P}
P_{E}=\mathbb{E}\left\{ \mathrm{p}_{\hat{k}}^{E2E}\right\} \leq\alpha\underset{\Gamma_{\hat{k}}}{\mathbb{E}}\left\{ Q\left(\sqrt{\beta\mu\Gamma_{\hat{k}}}\right)\right\} \leq\alpha\underset{\widetilde{\Gamma}_{\widetilde{k}}}{\mathbb{E}}\left\{ Q\left(\sqrt{\beta\mu\widetilde{\Gamma}_{\widetilde{k}}}\right)\right\}.
\end{equation}Then, we focus on the diversity gain and array power gain reflected by the upper bound in $(\ref{A_U_P})$. Using the method in \cite{GBG}, the upper bound is approximated as
\begin{equation}\label{A_U1_P}
P_{E}\leq\alpha\underset{\widetilde{\Gamma}_{\widetilde{k}}}{\mathbb{E}}\left\{ Q\left(\sqrt{\beta\mu\widetilde{\Gamma}_{\widetilde{k}}}\right)\right\} \thickapprox\alpha\left(G_{c}\mu\right)^{-G_{d}},
\end{equation}where
$G_{c}$ characterizes the array power gain, $G_{d}$ denotes the diversity order. In order to obtain the values of $G_{c}$ and $G_{d}$, the stochastic property of $\widetilde{\Gamma}_{\widetilde{k}}$ is investigated. Particularly, noticing that {\small{$$\widetilde{\Gamma}_{\widetilde{k}}=\underset{_{k=1,\ldots,K}}{\max}\sum_{l=1}^{L_{k}}\min\left\{ |h_{kl}|^{2},|g_{kl}|^{2}\right\},$$}}$\widetilde{\Gamma}_{\widetilde{k}}\mu$ can be treated as the received SNR by applying a two-fold diversity technique (detailed later) to a reference $1\times L$ SIMO P2P system, where the destination's $L$ antennas are divided into $K$ groups and the $k$-th group consists of $L_k$ antennas. The two-fold diversity technique is detailed as follows.
\begin{enumerate}
\item Firstly, for each value of $k$, ($k=1,\ldots K$), $L_{k}$ diversity branches are combined using the maximum ratio combining (MRC) techniques. These branches
are mutually independent, and the $l$-th diversity branch is characterized by $\min\left\{ |h_{kl}|^{2},|g_{kl}|^{2}\right\} $. By applying MRC, the aggregated channel gain is $\sqrt{\sum_{l=1}^{L_{k}}\min\left\{ |h_{kl}|^{2},|g_{kl}|^{2}\right\}}$, which is able to provide a diversity order of $G_{d,k}$ and an array power gain of $G_{c,k}$.
\item Secondly, the aggregated channel obtained from MRC is then treated as a diversity branch, and such $K$ diversity branches are combined using the
 selection combining (SC) technique. These branches
are mutually independent, and the $k$-th diversity branch is characterized by {\small{$\sqrt{\sum_{l=1}^{L_{k}}\min\left\{ |h_{kl}|^{2},|g_{kl}|^{2}\right\}}$}}. Finally, after using the two-fold diversity technique,
the resultant channel gain is $\widetilde{\Gamma}_{\widetilde{k}}$, which provides a diversity order of $G_{d}$ and an array power gain of $G_{c}$.
\end{enumerate}The two-fold diversity technique is termed as the MRC-SC scheme in our paper.
 Following the result of\cite{GBG} (Proposition 4), we have
\begin{equation}\label{AD_P}
\ensuremath{G_{d}}=\sum_{k=1}^{K}\ensuremath{G_{d,k}}=\sum_{k=1}^{K}L_{k}=L.
 \end{equation}The second equality in (\ref{AD_P}) is based on $\ensuremath{G_{d,k}}=L_{k}$, which can be obtained from classical performance analysis of the MRC technique. Again, relying on  \cite{GBG}, $G_{c}$ is expressed as
 \begin{equation}\label{AA_P}
 G_{c}=\left[\frac{2^{K-1}\pi^{\frac{K-1}{2}}Z\left(L+\frac{1}{2}\right)}{\prod_{k}G_{c,k}^{L_{k}}Z\left(L_{k}+\frac{1}{2}\right)}\right]^{-\frac{1}{L}},
 \end{equation}where $Z\left(t\right)=\int_{0}^{\infty}x^{t-1}\exp\left(-x\right)dx$, $G_{c,k}$ is the array power gain from combining $L_{k}$ diversity branches using MRC technique. According to the stochastic properties of $\min\left\{ |h_{kl}|^{2},|g_{kl}|^{2}\right\}$ ($l=1,\ldots L_{k}$) and \cite{GBG}, $G_{c,k}$ is expressed as
 \begin{equation}\label{AA1_P}
 G_{c,k}=\left[\frac{2^{L_{k}-1}\pi^{\frac{L_{k}-1}{2}}Z\left(\frac{1}{2}+L_{k}\right)}{Z\left(1+L_{k}\right)\left[\frac{\sqrt{\pi}}{2}\beta\right]^{L_{k}}}\right]^{\frac{-1}{L_{k}}}.
 \end{equation}Substituting (\ref{AA1_P}) into (\ref{AA_P}), the array power gain $ G_{c}$ is obtained.
 Notice that $\alpha\left(G_{c}\mu\right)^{-G_{d}}$ is an upper bound of $P_{E}$. Therefore,
in terms of achievable SER performance, using the proposed scheme in the two-way system is no worse than applying the MRC-SC scheme to the reference $1\times L$ SIMO P2P system, where the destination's $L$ antennas are divided into $K$ groups. The antenna partition in the latter reference SIMO system is identical to the antenna distribution among $K$ relays in the former two-way system. The analytical result of (\ref{AD_P}) reveals that the diversity order only depends on the total number of antennas. According to (\ref{AA_P}) and (\ref{AA1_P}), the array power gain of the proposed scheme comes from two contributing factors: the first one is the MRC-based processing detailed in 1), and the array power gain offered by this technique benefits from
 multiple antennas equipped in each relay. The second one is SC-based processing detailed in 2), the SC-based processing technique further strengthens the array power gain obtained from MRC-processing technique according to (\ref{AA_P}).
And the array power gain offered by this technique benefits from applying the proposed MaxMin RS scheme to $K$ multi-antenna relays. Since the proposed scheme at least obtains such two-fold array power gain, the performance of the proposed scheme outperforms the other existing schemes, which is corroborated by the simulation results provided in Section V given later.

\section{Numerical Results And Discussions}
In this section, we compare the SER of the proposed PR-MaxMin-RS scheme, the AF based scheme of \cite{AF-SER}, the MaxMin-AS scheme of \cite{eslamifarmax} and the O-TR scheme of \cite{zhou2010decode} by computer simulations. In all simulations, the channel coefficients and noises at each antenna are i.i.d. complex-valued Gaussian random variables with zero mean and unit variance (i.e., $\sigma^2=1$). In all given figures, the horizontal axis labeled by SNR (defined as $\frac{P_s}{\sigma^2}$) indicates the transmit power of the sources. The numerical results of the AF based scheme and the O-TR scheme are confirmed by the analytical results given in \cite{AF-SER} and \cite{zhou2010decode}. Different relay configurations are considered as well.
\subsection{Two Distributed Multi-Antenna Relays}
\begin{figure}
\begin{centering}
\includegraphics[height=6.9cm,width=8cm]{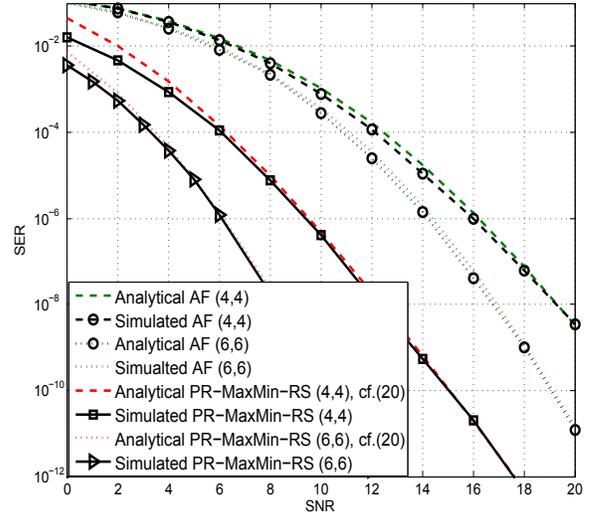}
\par\end{centering}
\caption{The end-to-end SER performance comparison between the proposed PR-MaxMin-RS scheme and the AF scheme, when BPSK modulation is employed.}
\end{figure}
In Fig. 3  and Fig. 4, the SER performance of the proposed PR-MaxMin-RS scheme, the O-TR scheme, and the AF scheme is compared using numerical simulations. Three configurations are considered: $L_{1}=L_{2}=4$,  $L_{1}=L_{2}=6$ and $L_{1}=L_{2}=8$, which are denoted as (4,4), (6,6), and (8,8), respectively. BPSK is employed. The value of $p$ is set to 2 such that $C_{2}p\min\left\{ \left|\mathbf{w}_{k}^{T}\mathbf{h}_{k}\right|^{2},\left|\mathbf{w}_{k}^{T}\mathbf{g}_{k}\right|^{2}\right\}
 $  equals to $\frac{\lambda_{LB}^{2}}{2}$. As shown in both figures, the three schemes considered achieve full diversity gain in all configurations. Furthermore, the proposed PR-MaxMin-RS scheme attains significantly higher array power gains than the AF scheme and the O-TR scheme do. This observation can be explained by the fact that the proposed PR-MaxMin-RS scheme fully utilizes all antennas in the selected relay to combine the signal power throughout the MA and BC stages. By contrast, the O-TR and the AF schemes only employ one selected antenna among the available $L_{1}+L_{2}$ antennas. This view is also supported by the observation from Fig. 3 that the performance gap is enlarged as $L_1$ and $L_2$ increase. In addition, the analytical upper bound of the SER in (\ref{P_E}) is also illustrated . It is seen that when the SNR becomes high, the upper bound converges to the numerical results.
\subsection{A Single Relay with Multiple Antennas}
In Fig. 5, we consider the centralized scenario where all relaying antennas are gathered in a single relay. The MaxMin-AS scheme and the transmission beamforming (TB) scheme of \cite{eslamifarmax} are compared with the proposed PR-MaxMin-RS scheme. According to Proposition 3, the PR preprocessing operation are removed. Hence, the corresponding overhead is avoided in the proposed PR-MaxMin-RS scheme. As shown in Fig.5, the TB scheme and the proposed PR-MaxMin-RS scheme achieve the similar performance, which is confirmed by the analytical result as well. Moreover, the numerical and analytical SER results of the MaxMin-AS scheme are also illustrated in Fig. 5. We can see that the proposed PR-MaxMin-RS scheme achieves higher array power gain than the MaxMin-AS scheme. This performance gap arises from the fact that the MaxMin-AS scheme only exploits the channel associated with one antenna in the BC stage, while the proposed PR-MaxMin-RS scheme combines the transmitted power of all antennas for the sources.

\begin{figure}
\begin{centering}
\includegraphics[height=7cm,width=8cm]{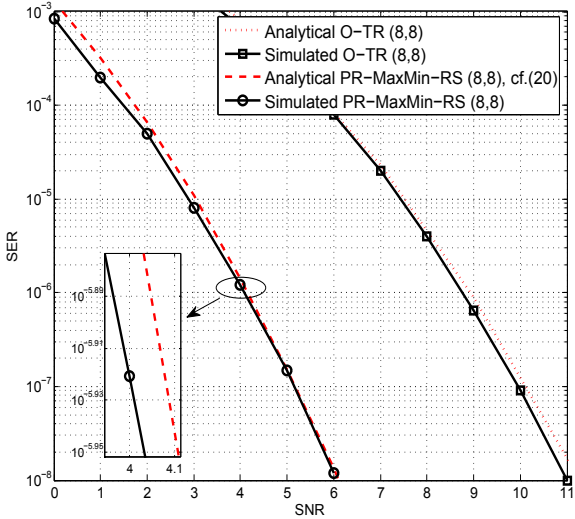}
\par\end{centering}
\caption{The end-to-end SER performance comparison between the proposed PR-MaxMin-RS scheme and the O-TR scheme, when BPSK modulation is employed.}
\end{figure}
\begin{figure}
\begin{centering}
\includegraphics[height=7cm,width=8cm]{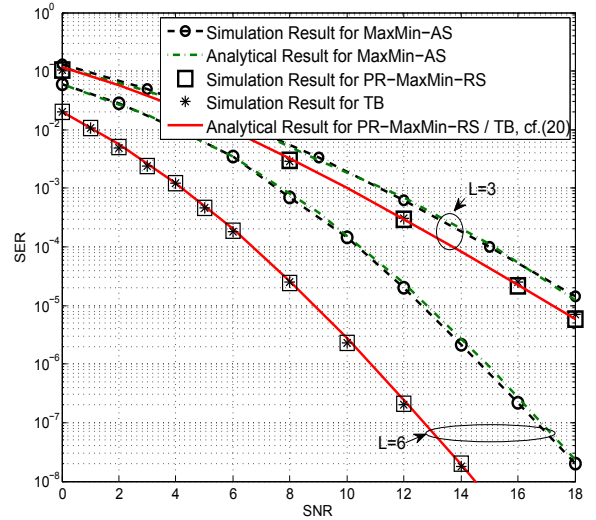}
\par\end{centering}
\caption{The end-to-end SER performance comparison between the proposed PR-MaxMin-RS scheme, the MaxMin-AS scheme and the TB scheme, for the cases of $K=1, L=3,6$, when BPSK modulation is employed.}
\end{figure}
\subsection{Multiple Distributed Single-Antenna Relays}
In Fig. 6, we compare the SER performance of the considered schemes in the configurations where there are 4 and 8 single-antenna relays. 4PAM modulation is employed and $p$ is set to 1. We consider the proposed PR-MaxMin-RS scheme, the AF based scheme and the straightforward MaxMin-RS scheme dispensing with any PR preprocessing. Two configurations, i.e., $L=4$ and $L=8$ are examined. It is observed that the proposed PR-MaxMin-RS scheme achieves the full diversity order of $L$. This is consistent with our analytical result given in Section IV. Nevertheless, it is obvious that the straightforward MaxMin-RS scheme fails to achieve the full diversity. This observation shows that the PR preprocessing operation is critical for the MaxMin-RS scheme to achieve full diversity. Furthermore, the PR-MaxMin-RS scheme also achieves better performance than the AF scheme. This can be explained by the fact that in the AF scheme, the selected relay forwards the desired signal together with the undesired noise, but the proposed PR-MaxMin-RS scheme only broadcasts the intended signal. In addition, we observe that the analytical SER given by (\ref{P_E}) converges to the numerical results in the high-SNR region.
\begin{figure}
\begin{centering}
\includegraphics[height=7cm,width=8cm]{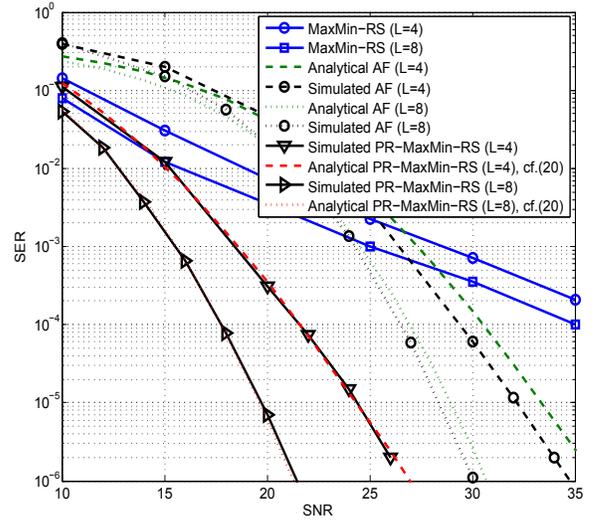}
\par\end{centering}
\caption{The end-to-end SER performance comparison between the proposed PR-MaxMin-RS scheme, the straightfoward MaxMin-RS scheme without PR preprocessing, and the AF scheme, when $K=L=4,8$, and 4PAM is employed.}
\end{figure}
\begin{figure}
\begin{centering}
\includegraphics[height=7cm,width=8cm]{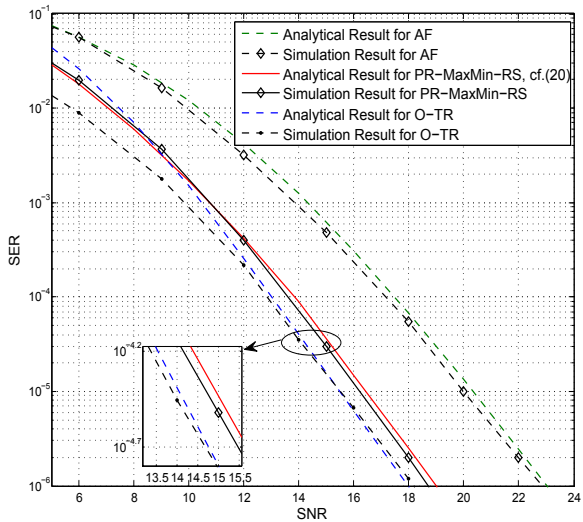}
\par\end{centering}
\caption{The end-to-end SER performance comparison between the proposed PR-MaxMin-RS scheme, the O-TR scheme, and the AF scheme, where $K=L=4$, and BPSK is employed.}
\end{figure}
In summary, although the proposed RS scheme optimizes the upper bound of the instantaneous SER, as shown in (\ref{P_bound}), rather than the actual SER,  the numerical results have shown that the average SER performance exhibits full diversity gain as well as better array power gain than existing schemes. The numerical results also corroborate our  analytical results, demonstrating that the upper bound of SER given by (\ref{P_bound}) is suitable and the tightness of the upper bound does not impose a negative impact on the full diversity gain and array power gain achieved.

Fig. 7 provides the SER comparison between the proposed PR-MaxMin-RS scheme, the O-TR scheme of \cite{zhou2010decode} and the AF scheme of \cite{AF-SER}, when BPSK modulation instead of 4PAM is employed. This figure shows that the proposed PR-MaxMin-RS scheme achieves better performance than the AF scheme, but is inferior to the O-TR scheme by no more than 1dB. We point out that the NCS is assumed to be accurately known by relays in the O-TR scheme. By contrast, in the proposed PR-MaxMin-RS scheme, the relays do not need to know the accurate NCS. Compared with the PR-MaxMin-RS scheme, the O-TR scheme assumes that more knowledge about the NCS is available to the relays. Therefore, under this condition it is reasonable for the O-TR to obtain better performance than the PR-MaxMin-RS scheme. Moreover, in the O-TR scheme all relays need to decode their received signals in the MA stage\footnote{Comparing the decoded result with the known NCS, each relay checks whether itself should be selected in the BC stage.}, by comparison, such requirement is removed in the PR-MaxMin-RS scheme. On the other hand, we acknowledge that our proposed scheme requires CSI feedback, while the O-TR scheme does not need. Consequently, our proposed scheme is not well suitable for the fast-varying fading scenario, where the channels only keep static during a small number of symbol durations. In principle, the CSI feedback may be employed frequently as the channel varies. However, the overhead cost for CSI feedback is no longer negligible in this scenario. It is important to relax or remove the CSI feedback constraint, which will be studied in our future work.

\subsection{Impact of Channel Estimation Error}
Finally, although our analysis focuses on an RS approach under the assumption of perfect channel knowledge, in Fig. 8 we also show the impact of imperfect
CSI encountered during the RS process. The channel estimation error is modeled as an complex-valued additive white Gaussian noise (AWGN) having a variance
$\delta^{2}$. This AWGN is added to the instantaneous channel coefficient and affects the RS performance \cite{ICEE}. More specifically, Fig. 8
shows the SER performance of the proposed PR-MaxMin-RS scheme under different levels of channel estimation error indicated by $\delta^{2}=0.1,0.5,0.01,0$. The simulation
employs QPSK modulation and two single-antenna relays. It is observed from Fig. 8 that when the channel estimation error is serious ($\delta^{2}=0.1,0.5$), the SER performance of our proposed scheme is significantly degraded, which also indicates a deteriorated achievable diversity gain. As the channel estimation
error increases, the SER performance degradation consistently becomes stronger, which yields a loss of the achievable diversity gain at high SNRs. On the other hand, when the channel estimation
error is alleviated to $\delta^{2}=0.01$, the achievable SER performance and the corresponding diversity gain of our proposed scheme are close to those of the perfect CSI scenario.
This observation reveals that our proposed scheme is sensitive to the serious channel estimation error. Therefore, a high-performance  channel-estimation algorithm is required to make the proposed scheme work effectively. Alternatively, it also implies that developing robust RS methods which are insensitive to channel estimation error is of great importance in the future work.

\begin{figure}
\begin{centering}
\includegraphics[height=7cm,width=8cm]{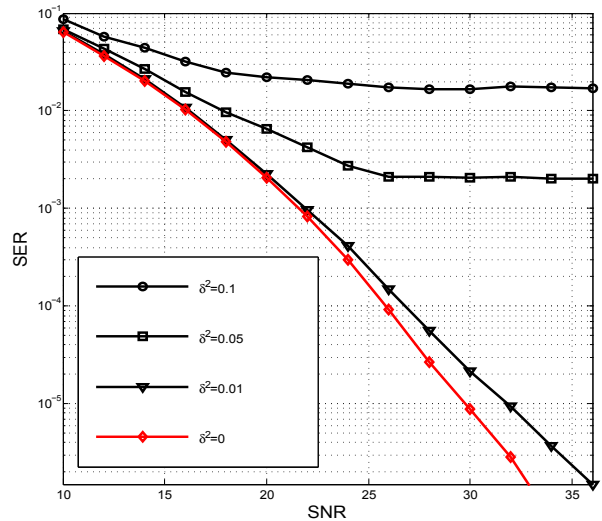}
\par\end{centering}
\caption{The end-to-end SER performance of PR-MaxMin-RS scheme under different levels of channel estimation error, when $K=L=2$, and QPSK is employed.}
\end{figure}

\section{Conclusions}
This paper investigated a family of PR strategies that facilitate the MaxMin-RS criterion to achieve full diversity in DF aided TWR systems. Specifically, when the sources transmit to the selected relay, they rotate the transmitted constellation symbols according to the phases of the selected bidirectional channels. When the relay broadcasts the decoded NCS, the PR strategy is employed again so that the NCS is rotated corresponding to the phases of the broadcast channel. By judiciously using the PR strategy in both of the the two aforementioned phases, the system eventually achieves full diversity in terms of the end-to-end SER.  Additionally, we show by both numerical and analytical results that the proposed PR-MaxMin-RS schemes outperforms other existing relaying schemes under a variety of relay configurations in the considered TWR system.

\appendices{}
\section{The proof of Proposition 1}
\begin{IEEEproof}
Depending on the value of $\left(d_{1},d_{2}\right)$, $\lambda^{2}\left(u_1d_1,u_2d_2\right)$ has two kinds of lower bounds. In the case of $d_{1}\neq0$ and $d_{2}\neq0$,
\begin{align}
\label{lowerbound2}
\nonumber &\lambda^{2}\left(u_1d_1,u_2d_2\right)=\|\mathbf{g}_{{{k}}}u_{2}d_{1}+\mathbf{h}_{{{k}}}u_{1}d_{2}\|^{2}\\ \nonumber&=\left\Vert \mathbf{g}_{{k}}d_{1}\right\Vert ^{2}+\left\Vert \mathbf{h}_{{k}}d_{2}\right\Vert ^{2}+2\Re\left\{ \exp\left(j\upsilon\right)\exp\left(j\varphi_{k}\right)\mathbf{g}_{{k}}^{H}\mathbf{h}_{{k}}d_{2}^{\text{*}}d_{1}\right\}  \\ \nonumber &=\left\Vert \mathbf{g}_{{k}}d_{1}\right\Vert ^{2}+\left\Vert \mathbf{h}_{{k}}d_{2}\right\Vert ^{2}+2\left|\mathbf{g}_{k}^{H}\mathbf{h}_{k}d_{2}^{\text{*}}d_{1}\right|\cos\left(\angle d_{1}-\angle d_{2}+\upsilon\right)\\ \nonumber & \geqslant2\left\Vert \mathbf{g}_{{k}}d_{1}\right\Vert \left\Vert \mathbf{h}_{{k}}d_{2}\right\Vert -2\left\Vert \mathbf{g}_{{k}}d_{1}\right\Vert \left\Vert \mathbf{h}_{{k}}d_{2}\right\Vert \left|\cos\left(\angle d_{1}-\angle d_{2}+\upsilon\right)\right|\\&\geqslant\Gamma_{k}d_{min}^{2}\underbrace{2\left(1-\left|\cos\left(\angle d_{1}-\angle d_{2}+\upsilon\right)\right|\right)}_{\rho\left(\upsilon, d_{1}, d_{2}\right)},
\end{align}where the third equality holds owing to by $\mathbf{g}_{{{k}}}^{H}\mathbf{h}_{{{k}}}=\left|\mathbf{g}_{{{k}}}^{H}\mathbf{h}_{{{k}}}\right|\exp\left(-j\varphi_{k}\right)$. Note that $\rho\left(\upsilon, d_1, d_2 \right)$ is a constant depending on $\upsilon$ for given $d_1$ and $d_2$.

On the other hand, for the cases $\left\{ d_{1}=0,d_{2}\neq0\right\} $ and $\left\{ d_{1}\neq0,d_{2}=0\right\}$, we have
\begin{equation}
\label{lowerbound3}
\lambda^{2}\left(u_1d_1,u_2d_2\right)\geqslant\Gamma_{k}d_{min}^{2}.
\end{equation}Combining (\ref{lowerbound2}) and (\ref{lowerbound3}), a general lower bound of $\lambda\left(u_1d_1,u_2d_2\right)$ is given by
\begin{align}
\label{lowerbound4}
\lambda\left(u_1d_{1},u_2d_{2}\right)&\geqslant\underset{\left(d_{1},d_{2}\right)\in D}{\min}\left\{ \sqrt{\Gamma_{k}d_{min}^{2}},\sqrt{\rho\left(d_{1},d_{2},\upsilon\right)\Gamma_{k}d_{min}^{2}}\right\} \\&=\sqrt{C_{3}\Gamma_{k}d_{min}^{2}},
\end{align}where the last equality holds due to $C_{3}\triangleq\min\left(1,\rho_{\min}\left(\upsilon\right)\right)$ and $\rho_{\min}\left(\upsilon\right)=\underset{\left(d_{1},d_{2}\right)\in D}{\min}\rho\left(d_{1},d_{2},\upsilon\right)$. Substituting (\ref{lowerbound4}) into (\ref{CP_SER}), (\ref{lowerbound1}) is obtained. Hence, the proof of Proposition 1 is established.
\end{IEEEproof}
\section{The proof of Proposition 2}
\begin{IEEEproof}
According to (\ref{GS}), we also have
\begin{equation}
\label{channel decompose}
\mathbf{h}_{{{k}}}=r_{11}\mathbf{q}_{1}^{T},\quad \mathbf{g}_{{k}}=r_{12}\mathbf{q}_{1}^{T}+r_{22}\mathbf{q}_{2}^{T}.
\end{equation}and
\begin{equation}
\label{norm}
||\mathbf{h}_{{{k}}}||^{2}=r_{11}^{2},||\mathbf{g}_{{k}}||^{2}=|r_{12}|^{2}+r_{22}^{2}.
\end{equation}Based on above properties of $\mathbf{h}_{{{k}}}$, $\mathbf{g}_{{{k}}}$ and $\mathbf{w}_{{k}}$, the impact of the PR beamforming scheme on $\mathrm{p}_{k,i}^{BC}$ is revealed as follows.
Firstly, ${\rm{S}}_1$ receives the signal $y_{BC,1}$ in form of (\ref{BCreceived}) as
\begin{align}
\nonumber y_{BC,1}&=\sqrt{P_r}\mathbf{h}_{{k}}^{T}\mathbf{w}_{{k}}{\hat{s}}_{NC}+n_{1}\\ \nonumber &=\sqrt{\frac{P_r}{2}}r_{11}\mathbf{q}_{1}\left(\exp\left(-j\varphi_{k}\right)\mathbf{q}_{1}^{H}+j\mathbf{q}_{2}^{H}\right){\hat{s}}_{NC}+n_{1}\\ &=\sqrt{\frac{P_r}{2}}r_{11}\exp\left(-j\varphi_{k}\right){\hat{s}}_{NC}+n_{1},
\end{align}where the second equality follows by using $\mathbf{h}_{{{k}}}=r_{11}\mathbf{q}_{1}^{T}$ given in (\ref{channel decompose}), and the last equality follows by employing $\mathbf{q}_{1}\mathbf{q}_{2}^{H}=0$. Hence, as shown in (\ref{PBC}) $\mathrm{p}_{k,1}^{BC}$ is given by
{\small{\begin{equation}
\label{mbcl1}
\mathrm{p}_{k,1}^{BC}=C_{1}Q\left(\sqrt{\frac{C_{2}p\mu}{2}|r_{11}|^{2}}\right)\overset{(a)}{=}C_{1}Q\left(\sqrt{\frac{C_{2}p\mu}{2}\left\Vert \mathbf{h}_{{k}}\right\Vert ^{2}}\right),
\end{equation}}}where the equality (a) follows by exploiting the fact that $||\mathbf{h}_{{{k}}}||^{2}=r_{11}^{2}$ as given by (\ref{norm}). Similarly,  the received signal of ${\rm{S}}_2$ becomes
\begin{align}
\nonumber y_{BC,2}&=\sqrt{P_r}\mathbf{g}_{{k}}^{T}\mathbf{w}_{{k}}{\hat{s}}_{NC}+n_{1}\\ \nonumber &=\sqrt{\frac{P_r}{2}}\left(r_{12}\mathbf{q}_{1}+r_{22}\mathbf{q}_{2}\right)\left(\exp\left(-j\varphi_{k}\right)\mathbf{q}_{1}^{H}+j\mathbf{q}_{2}^{H}\right){\hat{s}}_{NC}+n_{1}\\ \nonumber&=\sqrt{\frac{P_r}{2}}\left(r_{12}\exp\left(-j\varphi_{k}\right)+jr_{22}\right){\hat{s}}_{NC}+n_{1}\\&=\sqrt{\frac{P_r}{2}}\left(\left|r_{12}\right|+jr_{22}\right){\hat{s}}_{NC}+n_{1}
\end{align}where the second equality follows by using $\mathbf{g}_{{k}}=r_{12}\mathbf{q}_{1}^{T}+r_{22}\mathbf{q}_{2}^{T}$ as given in(\ref{channel decompose}), and the last equality follows by $\varphi_k=\angle\frac{\mathbf{h}_{{{k}}}^{H}\mathbf{g}_{{{k}}}}{|\mathbf{h}_{{{k}}}^{H}\mathbf{g}_{{{k}}}|}=\angle r_{12}$. Then, we have
{\small{\begin{align}
\label{mbcl2}
\mathrm{p}_{k,2}^{BC}&=C_{1}Q\left(\sqrt{\frac{C_{2}p\mu}{2}\left(\left|r_{12}\right|+jr_{22}\right)^{2}}\right)\\\nonumber&=C_{1}Q\left(\sqrt{\frac{C_{2}p\mu}{2}\left(|r_{12}|^{2}+r_{22}^{2}\right)}\right)\overset{}{=}C_{1}Q\left(\sqrt{\frac{C_{2}p\mu}{2}\left\Vert \mathbf{g}_{{k}}\right\Vert ^{2}}\right),
\end{align}}}where the second equality is based on the fact that $r_{22}$ is real-valued and the last equality results from $||\mathbf{g}_{{k}}||^{2}=|r_{12}|^{2}+|r_{22}|^{2}$ as given in (\ref{norm}).
Combining (\ref{mbcl1}), (\ref{mbcl2}) and (\ref{BClowerbound}), we obtain the general upper bound for $\mathrm{p}_{k,2}^{BC}$ as stated in (\ref{bcl}).
\end{IEEEproof}
\section{The proof of (\ref{e2eb})}
\begin{IEEEproof}
First, let us prove $\Gamma_{\hat{k}}\geqslant\xi$. Note that
\begin{equation}
\label{A1}
\Gamma_{\hat{k}}\text{=}\underset{_{k=1,\ldots,K}}{\max}\Gamma_{k}=\underset{_{k=1,\ldots,K}}{\max}\min\left\{ ||\mathbf{h}_{k}||^{2},||\mathbf{g}_{k}||^{2}\right\}.
\end{equation}For any $\tilde{k}\in\left\{ 1,\ldots K\right\}$, we have
{\small{\begin{equation}
\label{A2}
\Gamma_{\hat{k}}\geqslant\Gamma_{\tilde{k}}\geqslant\sum_{l=1}^{L_{\tilde{k}}}\min\left\{ |h_{\tilde{k}l}|^{2},|g_{\tilde{k}l}|^{2}\right\} \geqslant\min\left\{ |h_{\tilde{k}\hat{l}}|^{2},|g_{\tilde{k}\hat{l}}|^{2}\right\} =\xi,
\end{equation}}}
where the second inequality follows by invoking {\small{\begin{align}
\label{A3}
\nonumber \Gamma_{\tilde{k}}=\min\left\{ ||\mathbf{h}_{\tilde{k}}||^{2},||\mathbf{g}_{\tilde{k}}||^{2}\right\} &=\min\left\{ \sum_{l=1}^{L_{\tilde{k}}}|h_{\tilde{k}l}|^{2},\sum_{l=1}^{L_{\tilde{k}}}|g_{\tilde{k}l}|^{2}\right\} \\&\geqslant\sum_{l=1}^{L_{\tilde{k}}}\min\left\{ |h_{\tilde{k}l}|^{2},|g_{\tilde{k}l}|^{2}\right\}.
\end{align}}}Let
{\small{\begin{equation}
\label{virtual_MM}
\left(\tilde{k},\hat{l}\right)\triangleq\underset{_{k=1,\ldots,K,l=1,\ldots,L_{k}}}{\arg\max}\min\left\{ |h_{kl}|^{2},|g_{kl}|^{2}\right\}.
\end{equation}}}Then, (\ref{A3}) implies that $\Gamma_{\hat{k}}\geqslant\xi\triangleq\min\left\{ |h_{\tilde{k}\hat{l}}|^{2},|g_{\tilde{k}\hat{l}}|^{2}\right\}$. Next, we consider the stochastic property of $\xi$. According to (\ref{virtual_MM}), the $\hat{l}$th antenna in the $\tilde{k}$th relay is just the selection result of the MaxMin-AS scheme of \cite{eslamifarmax} applied to all the $L$ relaying antennas. It is given in \cite{eslamifarmax} that $\alpha\underset{\xi}{\mathbb{E}}\left\{Q\left(\sqrt{\beta\mu\xi}\right)\right\}\propto\frac{1}{\mu^{L}}$.
\end{IEEEproof}
\section{The proof of Proposition 5}
\begin{IEEEproof}
We rewrite (\ref{lowerbound2}) as
\begin{equation}
\label{re_lb}
\lambda^{2}\left(u_1d_1,u_2d_2\right)\geqslant\Gamma_{k}|d_{min}|^{2}\underbrace{2\left(1-|\cos\left(\angle d_{1}-\angle d_{2}+\upsilon\right)|\right)}_{\rho\left(\upsilon, d_1, d_2 \right)},
\end{equation}
where $\angle d_{1}$ and $\angle d_{2}$ equal either $0$ or $\pi$ when MPAM is employed. If $\upsilon$ is set to $\frac{\pi}{2}$, $|\cos\left(\angle d_{1}-\angle d_{2}+\upsilon\right)|$ is $0$. Correspondingly, $\rho\left(\upsilon, d_1, d_2 \right)$ always approaches its maximum value of $2$. The lower bound of $\lambda\left(u_1d_1,u_2d_2\right)$ in (\ref{lowerbound2}) is thus maximized. Furthermore, by substituting $\upsilon=\frac{\pi}{2}$ into (\ref{lowerbound2}), the average SER in the MA stage becomes
{\small{\begin{equation}
\label{mpam}
{P_{MA}}=\underset{\left(d_{1},d_{2}\right)\in D}{\sum}\frac{1}{M^2}\underset{\mathbf{g}_{{\hat{k}}},\mathbf{h}_{{\hat{k}}}}{\mathbb{E}}\left\{ Q\left(\sqrt{\frac{\mu}{2}\left(\left\Vert \mathbf{g}_{{\hat{k}}}d_{1}\right\Vert ^{2}+\left\Vert \mathbf{h}_{{\hat{k}}}d_{2}\right\Vert ^{2}\right)}\right)\right\}.
\end{equation}}}

We first reshape the term {\small{$\underset{\mathbf{g}_{{\hat{k}}},\mathbf{h}_{{\hat{k}}}}{\mathbb{E}}\left\{
 Q\left(\sqrt{\frac{\mu}{2}\left(\left\Vert \mathbf{g}_{{\hat{k}}}d_{2}\right\Vert ^{2}+\left\Vert \mathbf{h}_{{\hat{k}}}d_{1}\right\Vert ^{2}\right)}\right) \right\}$}} in (\ref{mpam}) as follows:
{\small{\begin{align}
\label{temp1}
\nonumber &\underset{\mathbf{g}_{{\hat{k}}},\mathbf{h}_{{\hat{k}}}}{\mathbb{E}}\left\{
 Q\left(\sqrt{\frac{\mu}{2}\left(\left\Vert \mathbf{g}_{{\hat{k}}}d_{2}\right\Vert ^{2}+\left\Vert \mathbf{h}_{{\hat{k}}}d_{1}\right\Vert ^{2}\right)}\right) \right\}	\\ \nonumber &=\underset{\mathbf{g}_{{\hat{k}}},\mathbf{h}_{{\hat{k}}}}{\mathbb{E}}\left\{\frac{1}{\pi}\int_{0}^{\frac{\pi}{2}}\exp\left(-\frac{\frac{\mu}{2}\left(\left\Vert \mathbf{g}_{{\hat{k}}}d_{2}\right\Vert ^{2}+\left\Vert \mathbf{h}_{{\hat{k}}}d_{1}\right\Vert ^{2}\right)}{2\sin^{2}\theta}\right)d\theta \right\}\\
	&=\frac{1}{\pi}\int_{0}^{\frac{\pi}{2}}\psi_{\gamma_{1}}\left(\frac{\left\Vert d_{1}\right\Vert ^{2}}{4\sin^{2}\theta}\right)\psi_{\gamma_{2}}\left(\frac{\left\Vert d_{2}\right\Vert ^{2}}{4\sin^{2}\theta}\right)d\theta
\end{align}}}where $\psi_{\gamma_{1}}\left(t\right)$ and $\psi_{\gamma_{2}}\left(t\right)$ denote the MGF of $\gamma_{1}=\mu \Vert \mathbf{h}_{{\hat{k}}}\Vert ^{2}$ and $\gamma_{2}=\mu \Vert \mathbf{g}_{{\hat{k}}}\Vert ^{2}$, respectively. The second equality is established on the fact that $\mathbf{g}_{{\hat{k}}}$ and $\mathbf{h}_{{\hat{k}}}$ are independently conditioned on $\hat{k}$ and that $\mathbf{g}_{{\hat{k}}}$ and $\mathbf{h}_{{\hat{k}}}$ have the same distribution,  $\psi_{\gamma_{1}}\left(t\right)=\psi_{\gamma_{2}}\left(t\right)$. On the other hand,
in this case, the selected relay has only one antenna. Hence, $\hat{s}_{NC}$ is broadcasted straightforwardly. As a result, $P_{BC,1}$ is given by
{\small{\begin{align}
\label{bc_pam}
\nonumber P_{BC,1}&=\underset{\mathbf{h}_{{\hat{k}}}}{\mathbb{E}}\left\{\frac{2\left(M-1\right)}{M\pi}Q\left(\sqrt{\frac{6\left\Vert \mathbf{h}_{{\hat{k}}}\right\Vert ^{2}p\mu}{M^{2}-1}}\right)\right\}\\ &=\frac{2\left(M-1\right)}{M\pi}\int_{0}^{\frac{\pi}{2}}\psi_{\gamma_{1}}\left(\frac{pg_{MPAM}}{\sin^{2}\theta}\right)d\theta.
\end{align}}}
In order to get $P_{MA}$ and $P_{BC,1}$,
we focus on $\psi_{\gamma_{1}}\left(t\right)$ in what follows. First, we compute the cdf (Cumulative Distribution Function) of
$\Vert \mathbf{h}_{{\hat{k}}}\Vert ^2$ as
{\small{\begin{align}
\label{F_h1}
\nonumber F\left(z\right)&=\Pr\left\{\Vert \mathbf{h}_{{\hat{k}}}\Vert ^2\leqslant z\right\} \\\nonumber &=\Pr\left\{ \Vert \mathbf{h}_{{\hat{k}}}\Vert ^2\leqslant z\left|\Vert \mathbf{h}_{{\hat{k}}}\Vert ^2\leqslant\Vert \mathbf{g}_{{\hat{k}}}\Vert ^2\right.\right\} \Pr\left\{\Vert \mathbf{h}_{{\hat{k}}}\Vert ^2\leqslant\Vert \mathbf{g}_{{\hat{k}}}\Vert ^2\right\} \\ \nonumber
	&+\Pr\left\{\Vert \mathbf{h}_{{\hat{k}}}\Vert ^2\leqslant z\left|\Vert \mathbf{h}_{{\hat{k}}}\Vert ^2\geqslant\Vert \mathbf{g}_{{\hat{k}}}\Vert ^2\right.\right\} \Pr\left\{ \Vert \mathbf{h}_{{\hat{k}}}\Vert ^2\geqslant\Vert \mathbf{g}_{{\hat{k}}}\Vert ^2\right\} \\ \nonumber
	&=\frac{1}{2}\Pr\left\{ \Vert \mathbf{h}_{{\hat{k}}}\Vert ^2\leqslant z\left|\Vert \mathbf{h}_{{\hat{k}}}\Vert ^2\leqslant\Vert \mathbf{g}_{{\hat{k}}}\Vert ^2\right.\right\} \\&+\frac{1}{2}\Pr\left\{ \Vert \mathbf{h}_{{\hat{k}}}\Vert ^2\leqslant z\left|\Vert \mathbf{h}_{{\hat{k}}}\Vert ^2\geqslant \Vert \mathbf{g}_{{\hat{k}}}\Vert ^2\right.\right\}.
 \end{align}}}The last equality in (\ref{F_h1}) results from the fact that $\Vert \mathbf{h}_{{\hat{k}}}\Vert ^2$ and $\Vert \mathbf{g}_{{\hat{k}}}\Vert ^2$ are i.i.d conditioned on $\hat{k}$, which implies that $\Pr\left\{\Vert \mathbf{h}_{{\hat{k}}}\Vert ^2\leqslant\Vert \mathbf{g}_{{\hat{k}}}\Vert ^2\right\} =\Pr\left\{ \Vert \mathbf{h}_{{\hat{k}}}\Vert ^2\geqslant\Vert \mathbf{g}_{{\hat{k}}}\Vert ^2\right\} =\frac{1}{2}$. Furthermore, it is noted that {\small{$\Pr\left\{ \Vert\mathbf{h}_{{\hat{k}}}\Vert^{2}\leqslant z\left|\Vert\mathbf{h}_{{\hat{k}}}\Vert^{2}\leqslant\Vert\mathbf{g}_{{\hat{k}}}\Vert^{2}\right.\right\} =\Pr\left\{ \min\left\{ \Vert\mathbf{h}_{{\hat{k}}}\Vert^{2},\Vert\mathbf{g}_{{\hat{k}}}\Vert^{2}\right\} \leqslant z\right\}$}}, {\small{$\Pr\left\{ \Vert\mathbf{h}_{{\hat{k}}}\Vert^{2}\leqslant z\left|\Vert\mathbf{h}_{{\hat{k}}}\Vert^{2}\geqslant\Vert\mathbf{g}_{{\hat{k}}}\Vert^{2}\right.\right\} =\Pr\left\{ \max\left\{ \Vert\mathbf{h}_{{\hat{k}}}\Vert^{2},\Vert\mathbf{g}_{{\hat{k}}}\Vert^{2}\right\} \leqslant z\right\}$}}. Therefore, (\ref{F_h1}) becomes
\begin{equation}
\label{F_h2}
F\left(z\right)=\frac{1}{2}F_{min}\left(z\right)+\frac{1}{2}F_{max}\left(z\right),
\end{equation}where $F_{min}\left(z\right)$ is the cdf of $\min\left\{ \Vert\mathbf{h}_{{\hat{k}}}\Vert^{2},\Vert\mathbf{g}_{{\hat{k}}}\Vert^{2}\right\} $, and $F_{max}\left(z\right)$ is the cdf of $\max\left\{ \Vert\mathbf{h}_{{\hat{k}}}\Vert^{2},\Vert\mathbf{g}_{{\hat{k}}}\Vert^{2}\right\} $. From \cite{tse2005fundamentals}, we have
\begin{equation}
\label{F_min}
F_{min}\left(z\right)\text{=}\left(1-\left(\exp\left(-z\right)\sum_{l=0}^{\mathcal{L}-1}\frac{1}{l!}z^{l}\right)^{2}\right)^{K}.
\end{equation}On the other hand, $F_{max}\left(z\right)=\Pr\left\{ \max\left\{ \Vert\mathbf{h}_{{\hat{k}}}\Vert^{2},\Vert\mathbf{g}_{{\hat{k}}}\Vert^{2}\right\} \leqslant z\right\} .
$ In order to get $F_{max}\left(z\right)$, we decompose the event $\left\{ \max\left\{ \Vert\mathbf{h}_{{\hat{k}}}\Vert^{2},\Vert\mathbf{g}_{{\hat{k}}}\Vert^{2}\right\} \leqslant z\right\} $ as follows:
{\small{\begin{align}
&\nonumber \left\{ \max\left\{ \Vert\mathbf{h}_{{\hat{k}}}\Vert^{2},\Vert\mathbf{g}_{{\hat{k}}}\Vert^{2}\right\} \leqslant z\right\} \\\nonumber &=\underset{k=1,\ldots,K}{\cup}\left\{ {\rm{the}}\, k{\rm{th}}\, {\rm{relay}}\, {\rm{is}}\, {\rm{selected}},\max\left\{ \Vert\mathbf{h}_{{k}}\Vert^{2},\Vert\mathbf{g}_{{k}}\Vert^{2}\right\} \leqslant z\right\}\\ &=\underset{k=1,\ldots,K}{\cup}\left\{ u_{k}\leqslant\max\left\{ \Vert\mathbf{h}_{{{k}}}\Vert^{2},\Vert\mathbf{g}_{{{k}}}\Vert^{2}\right\}  ,\max\left\{ \Vert\mathbf{h}_{{{k}}}\Vert^{2},\Vert\mathbf{g}_{{{k}}}\Vert^{2}\right\}  \leqslant z\right\}
\end{align}}}where $u_{k}=\underset{k'=1,\ldots,K,k'\neq k}{\max}\min\left\{ \Vert\mathbf{h}_{{k'}}\Vert^{2},\Vert\mathbf{g}_{{k'}}\Vert^{2}\right\}$. Then,
{\small{\begin{align}
\label{Fz}
\nonumber &F_{max}\left(z\right)=\Pr\left\{ \max\left\{ \Vert\mathbf{h}_{{\hat{k}}}\Vert^{2},\Vert\mathbf{g}_{{\hat{k}}}\Vert^{2}\right\} \leqslant z\right\} \\\nonumber&=\sum_{k=1}^{K}\Pr\left\{ u_{k}\leqslant\max\left\{ \Vert\mathbf{h}_{{{k}}}\Vert^{2},\Vert\mathbf{g}_{{{k}}}\Vert^{2}\right\}  ,\max\left\{ \Vert\mathbf{h}_{{{k}}}\Vert^{2},\Vert\mathbf{g}_{{{k}}}\Vert^{2}\right\}  \leqslant z\right\} \\&\nonumber=\sum_{k=1}^{K}\int_{0}^{z}\Pr\left\{ u_{k}\leqslant \Vert\mathbf{h}_{{k}}\Vert^{2}\leqslant z,u_{k}\leqslant\Vert\mathbf{g}_{{k}}\Vert^{2}\leqslant z|u_{k}\right\} f\left(u_{k}\right)du_{k}\\&=K\int_{0}^{z}\left(\exp\left(-z\right)\sum_{l=0}^{\mathcal{L}-1}\frac{1}{l!}z^{l}-\exp\left(-u\right)\sum_{l=0}^{\mathcal{L}-1}\frac{1}{l!}u^{l}\right)^{2}f\left(u\right)du,
\end{align}}}where the last equality is based on the fact that $\Vert\mathbf{h}_{{\hat{k}}}\Vert^{2}$ and $\Vert\mathbf{g}_{{\hat{k}}}\Vert^{2}$ independently follow Gamma distribution conditioned on $k$. Due to the symmetry, the pdf (Probability Distribution Function) of $u_k$ is identical for all $k$, we thus remove the subscript of $f\left(u_{k}\right)$ to arrive at the last equality.
Note that $u_k$ can be interpreted as the maximum of $\min\left\{ \Vert\mathbf{h}_{{k}}\Vert^{2},\Vert\mathbf{g}_{{k}}\Vert^{2}\right\} $ among  $K-1$ values. The pdf of  $u_k$ can be given by the derivative of $F_{min}\left(z\right)$ while replacing $K$ with $K-1$. Based on these intermediate results by a few much trivial and tedious calculations,
the expression of $\psi_{\gamma_{1}}\left(t\right)$ in (\ref{mgf}) is obtained. With this expression of $\psi_{\gamma_{1}}\left(t\right)$, we complete the proof of Proposition 5.
\end{IEEEproof}

\bibliographystyle{IEEEtran}



\begin{IEEEbiography}[{\includegraphics[width=1in,height=1.25in,clip,keepaspectratio]{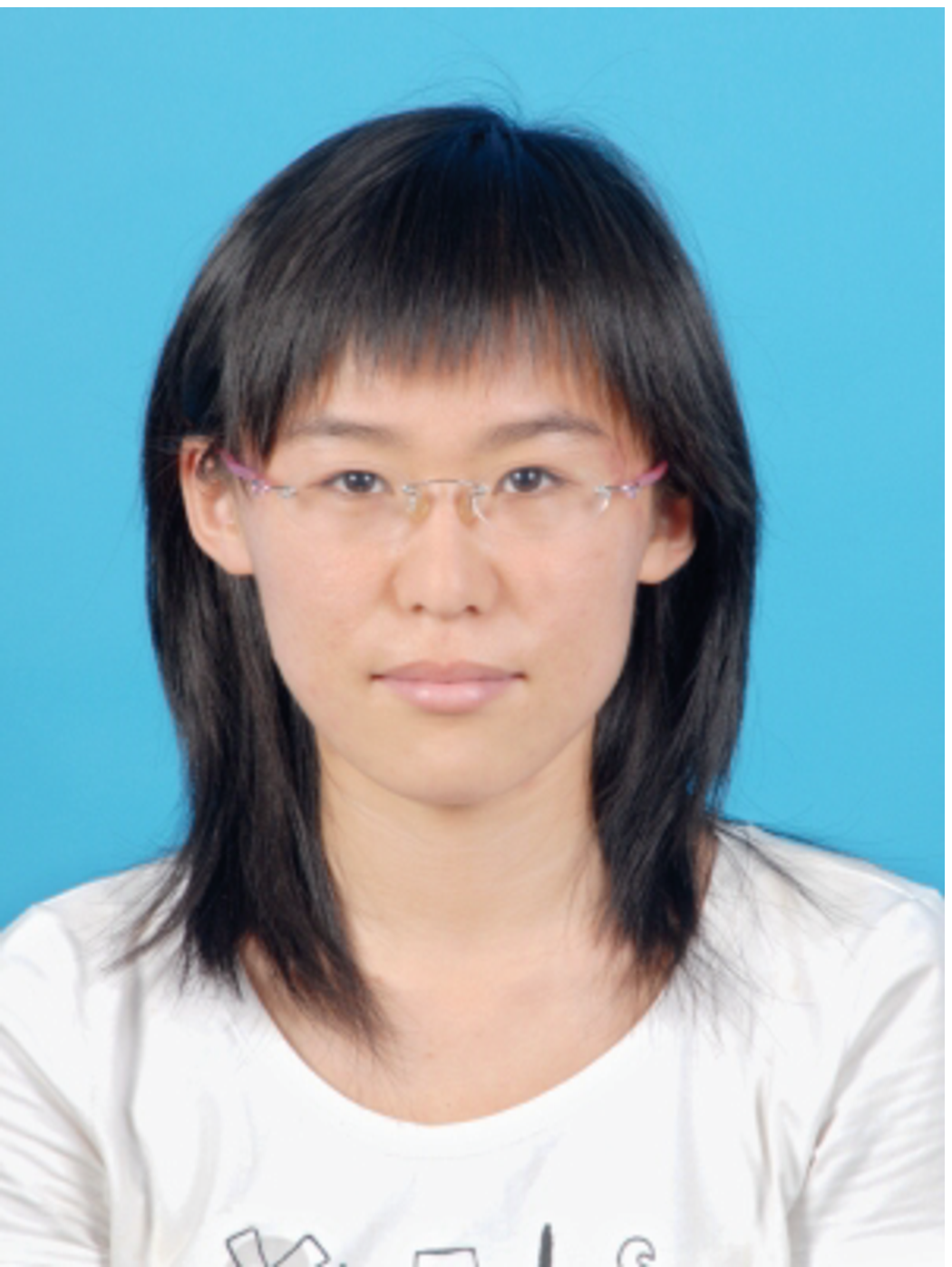}}]{Ruohan Cao} received her B.Eng. degree in 2009 from Shandong University of Science and Technology (SDUST), Qingdao, China. She received the Ph.D. degree in 2014
form Beijing University of Posts and Telecommunications (BUPT), Beijing, China. From November 2012 to August 2014, she also served as a research assistant for the Department of Electrical and Computer Engineering at University of Florida, supported by the China Scholarship Council.
She is now with the Institute of Information Photonics and Optical Communications, BUPT, as a Postdoc.
Her research interests include physical-layer network coding, multiuser multiple-input-multiple-output systems and physical-layer security.
\end{IEEEbiography}
\begin{IEEEbiography}[{\includegraphics[width=1in,height=1.25in,clip,keepaspectratio]{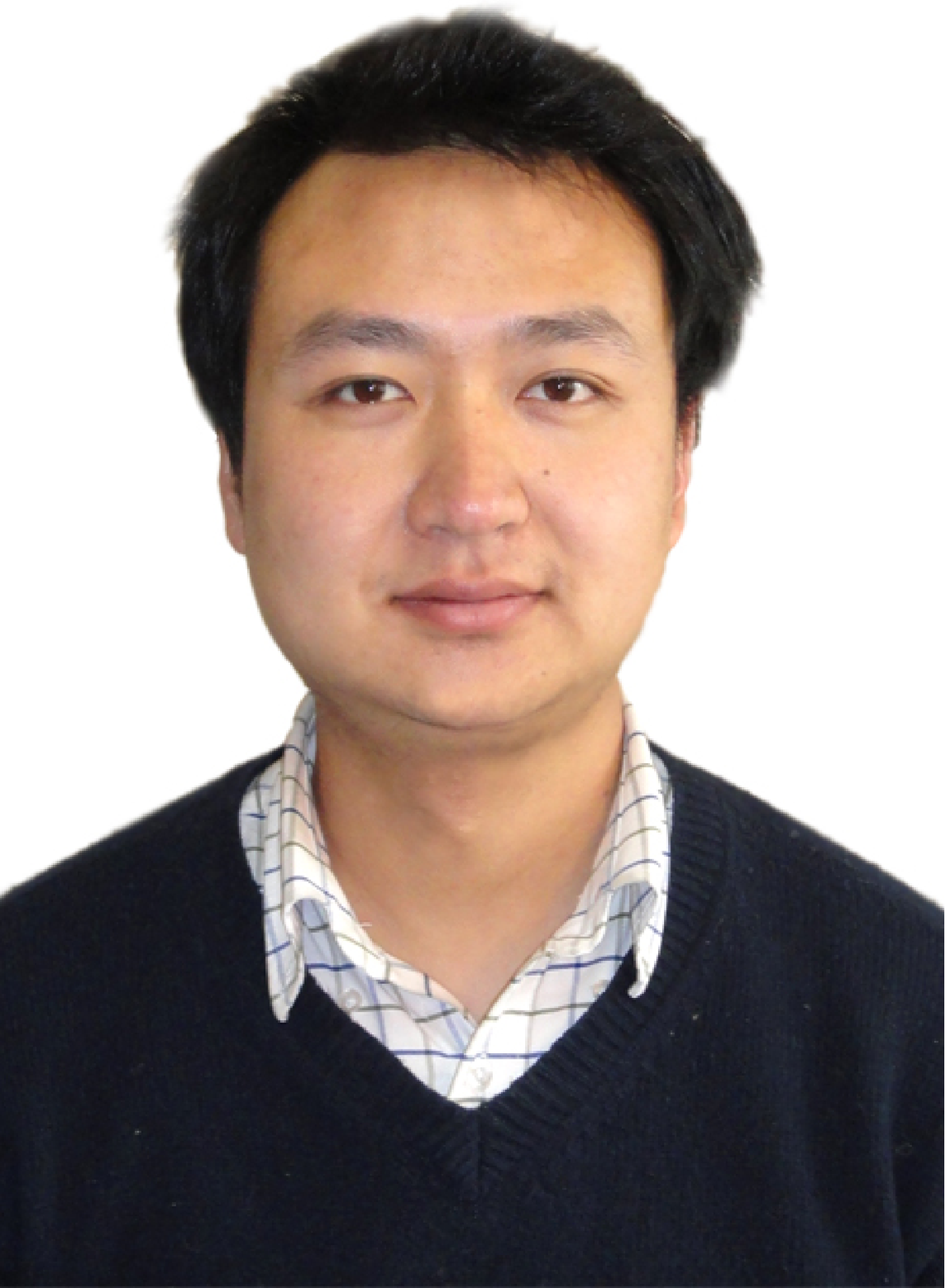}}]{Hui Gao} (S'10-M'13) received the B. Eng. degree in information engineering and the Ph.D. degree in signal and information processing from Beijing University of Posts and Telecommunications (BUPT), Beijing, China, in July 2007 and July 2012, respectively. From May 2009 to June 2012, he also served as a Research Assistant for the Wireless and Mobile Communications Technology R$\&$D Center, Tsinghua University, Beijing, China. From April 2012 to June 2012, he visited Singapore University of Technology and Design (SUTD), Singapore, as a Research Assistant. From July 2012 to February 2014, he was a Postdoc Researcher with SUTD. He is now with the School of Information and Communication Engineering, BUPT, as an Assistant Professor.
His research interests include massive MIMO systems, cooperative communications, ultra-wideband wireless communications.
\end{IEEEbiography}
\begin{IEEEbiography}[{\includegraphics[width=1in,height=1.25in,clip,keepaspectratio]{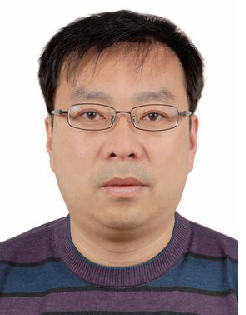}}]{Tiejun Lv }(M'08-SM'12) received the M.S. and Ph.D. degrees in electronic engineering from the University of Electronic Science and Technology of China (UESTC), Chengdu, China, in 1997 and 2000, respectively. From January 2001 to December 2002, he was a Postdoctoral Fellow with Tsinghua University, Beijing, China. From September 2008 to March 2009, he was a Visiting Professor with the Department of Electrical Engineering, Stanford University, Stanford, CA, USA. He is currently a Full Professor with the School of Information and Communication Engineering, Beijing University of Posts and Telecommunications (BUPT).
He is the author of more than 200 published technical papers on  the physical layer of wireless mobile communications.
His current research interests include signal processing, communications theory and networking.
He was the recipient of the Program for New Century Excellent Talents in University Award from the Ministry of Education, China, in 2006.
\end{IEEEbiography}
\begin{IEEEbiography}[{\includegraphics[width=1in,height=1.25in,clip,keepaspectratio]{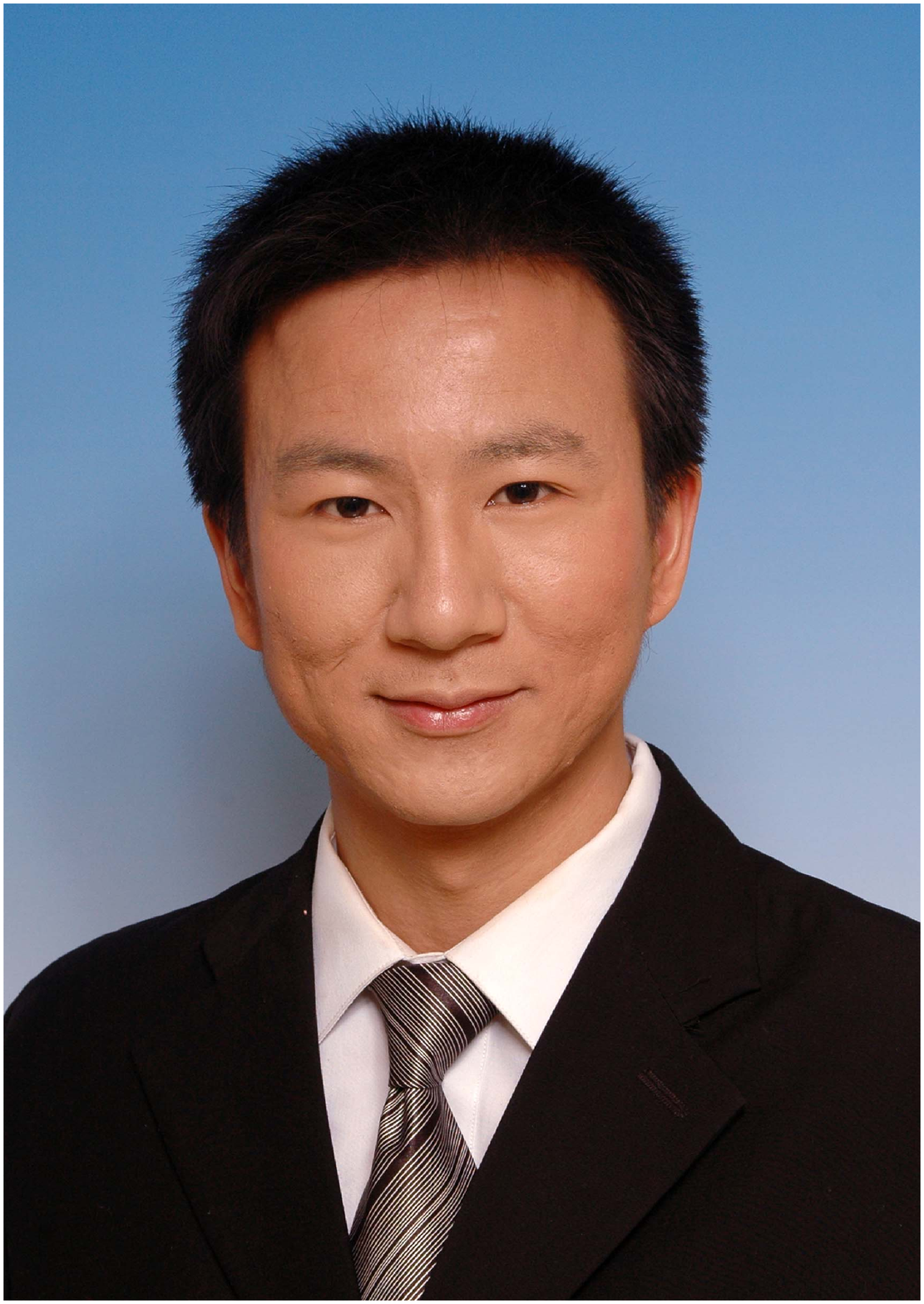}}]{Shaoshi Yang} (S'09-M'13) received the B.Eng. Degree in Information Engineering
from Beijing University of Posts and Telecommunications (BUPT), China, in 2006, the first Ph.D. Degree in Electronics and Electrical Engineering from University of Southampton, U.K., in 2013, and a second Ph.D. Degree in Signal and Information Processing from BUPT in 2014. Since 2013 he has been a Postdoctoral Research Fellow in University of Southampton, U.K, and from 2008 to 2009, he was an Intern Research Fellow with the Intel Labs China, Beijing, where he focused on
Channel Quality Indicator Channel design for mobile WiMAX (802.16 m). His research interests include MIMO signal processing, green radio, heterogeneous networks, cross-layer interference management, convex optimization and its applications. He has published
in excess of 30 research papers on IEEE journals and conferences.

Shaoshi has received a number of academic and research awards, including the PMC-Sierra Telecommunications Technology Scholarship at BUPT, the Electronics and Computer Science (ECS) Scholarship of University of Southampton and the Best PhD Thesis Award of BUPT. He serves as a TPC member of a number of IEEE conferences and journals, including \textit{IEEE ICC, PIMRC, ICCVE, HPCC} and \textit{IEEE Journal on Selected Areas in Communications}. He is also a Junior Member of the Isaac Newton Institute for Mathematical Sciences, Cambridge University, UK. (https://
sites.google.com/site/shaoshiyang/)
\end{IEEEbiography}
\begin{IEEEbiography}[{\includegraphics[width=1in,height=1.25in,clip,keepaspectratio]{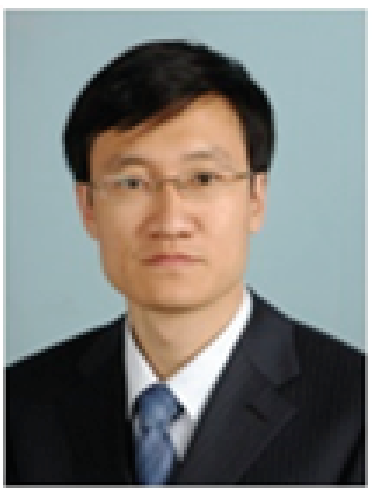}}]{Shanguo Huang} (M'09) received the Ph.D. degree from Beijing University of Posts and Telecommunications, Beijing, P. R. China, in 2006. He is currently a professor in the State Key Laboratory of Information Photonics and Optical Communications (IPOC), and vice dean in School of Electronic Engineering, in BUPT, P. R. China. He has been actively undertaking several national projects, published 3 books and more than 150 journals and refereed conferences, and authorized 14 patents. He was awarded the Beijing Higher Education Young Elite Teacher, the Beijing Nova Program, and the Program for New Century Excellent Talents in University from the Ministry of Education, in 2011-2013, respectively. His current research interests include the networks designing, planning, the traffic control and resource allocations, especially network routing algorithms and performance analysis.
\end{IEEEbiography}
\end{document}